\documentclass[epjST]{svjour}
 \usepackage{graphicx}% Include figure files
\usepackage{dcolumn}% Align table columns on decimal point
\usepackage{bm}% bold math
\usepackage{lipsum}
\usepackage{adjustbox}
\usepackage{multirow}
\usepackage{amsmath}
\usepackage{amssymb}
\usepackage{enumerate}
\graphicspath{{./eps/}}
\begin{document}
\title{Deep inelastic (anti)neutrino-nucleus scattering}
% \subtitle{Do you have a subtitle?\\ If so, write it here}
\author{V. Ansari$^{1}$, M. Sajjad Athar$^{1}$, H. Haider$^{1}$\thanks{email: huma.haider8@gmail.com}, I. Ruiz Simo$^{2}$, S. K. Singh$^{1}$ \and F. Zaidi$^{1}$}
\institute{$^{1}$Department of Physics, Aligarh Muslim University, Aligarh-202002, India\\
$^{2}$Departamento de F\'{\i}sica At\'omica, Molecular y Nuclear,
and Instituto de F\'{\i}sica Te\'orica y Computacional Carlos I,
Universidad de Granada, Granada 18071, Spain}
 \abstract{
The present status of the field theoretical model studies of the deep inelastic scattering induced by (anti)neutrino on the nuclear targets in a wide 
range of Bjorken variable $x$ and four momentum transfer square $Q^2$, has been reviewed~\cite{Haider:2011qs,Haider:2012nf,Haider:2016zrk,Zaidi:2019mfd,Zaidi:2019asc,Ansari:2020xne}.
 The effect of the nonperturbative corrections such as target mass correction and higher twist effects, perturbative evolution of the  parton densities, nuclear medium modifications in the nucleon structure functions, nuclear isoscalarity corrections on the weak nuclear structure functions have been discussed. These structure functions have been used to obtain the differential scattering cross sections. The various nuclear medium effects like the Fermi motion, binding energy, nucleon correlations, mesonic contributions, shadowing and antishadowing corrections relevant in the different regions of $x$ and $Q^2$ have been discussed. The numerical results for the structure functions and the cross sections are compared with some of the available experimental data including the recent
results from MINERvA. The predictions  are made in argon nuclear target which is planned to be used as a target material in  DUNE at the Fermilab.
 } 
\maketitle
\section{Introduction}
\label{intro}
In recent years the need for a better understanding of neutrino interaction cross section with nucleon and nuclear targets in the few GeV
  region of (anti)neutrino energies has been emphasized in order to reduce the systematic uncertainty in the analysis of neutrino 
 oscillation parameters of the Pontecorvo-Maki-Nakagawa-Sakata matrix (PMNS matrix). The lack in the understanding of (anti)neutrino-nucleon/nucleus cross section adds to 25-30$\%$ uncertainty to the systematics and a considerable reduction in this uncertainty is required 
 for a precise measurement of PMNS matrix or in the study of CP violation in the leptonic sector or in the determination of the neutrino mass
 hierarchy (normal or inverted). In the few GeV region of (anti)neutrino 
 energies the contribution to the total (anti)neutrino cross section comes from the quasielastic, inelastic as well as the deep inelastic scattering processes 
 on the nuclear targets. In this review, we shall focus on the understanding of nuclear medium effects in the weak interaction
 induced deep inelastic scattering(DIS) processes in a theoretical model using field theory. For the details on the current status readers are referred to Ref.\cite{SajjadAthar:2020nvy}.
 
There is intrinsic interest to understand DIS processes induced by the charged leptons and (anti)neutrinos on nucleons
and nuclear targets as they are important tools to study the quark parton structure of the free nucleons and the nucleons when they
are bound in a nucleus. The observation of EMC (European Muon Collaboration) effect in the DIS of muon from iron target
showed that the quark parton
distributions of free nucleons are considerably modified when they are bound in a nucleus which was later confirmed in several other experiments 
by using electron, muon, (anti)neutrino and other particle beams on the various nuclear targets. The DIS experiments done with lepton beams make measurements of the DIS cross sections which are expressed in
terms of the nucleon structure functions. These structure functions are determined by making Rosenbluth-like separations of the measured cross section. 
The electromagnetic DIS cross section induced by the charged leptons are generally expressed in terms of the two structure functions $F_{iN}^{\text{EM}}(x,Q^2 )$ $(i=1,2)$, 
which are functions of the Bjorken scaling variable  $x=\frac{Q^2}{2 M_N \nu}$ (0 $\le \; x \; \le \;$1); and $Q^2$ (where $Q^2 \ge$ 0, is the four momentum transfer square and
 $\nu$ is the energy transferred to the target i.e. $\nu=E_\nu-E_l$, $E_\nu(E_l)$ is the energy of the incoming(outgoing) lepton, and  $M_N$ is the nucleon mass). For the weak DIS process induced by 
 (anti)neutrinos the cross sections 
are given in terms of three structure functions $F_{iN}^{\text{Weak}}(x,Q^2 )$ $(i=1,2,3)$, in the limit of the vanishing lepton mass. 
 However, in the case of DIS induced by the weak charged  current of $\nu_\mu$ and $\nu_\tau$, where the lepton mass in the final state could be non-negligible as compared to $Q^2$ in some kinematic regions, then two additional structure functions $F_{4N}^{Weak}(x,Q^2)$ and $F_{5N}^{Weak}(x,Q^2)$ also contribute to the cross sections. In the exact limit of Bjorken scaling i.e. $\nu \rightarrow \infty$, $Q^2 \rightarrow \infty$, such that
$x=\frac{Q^2}{2M_N\nu}$,  remains fixed, the structure functions scale and become functions of only one variable $x$.  These structure functions are not all independent when calculated in the quark-parton model and satisfy certain relations given by Callan-Gross~\cite{Callan:1969uq} and Albright-Jarlskog~\cite{Albright:1974ts}. Consequently, the electromagnetic cross sections are given in terms of only one structure function chosen to be $F_{2N}^{EM}(x,Q^2)$ while the weak cross sections are given in terms of two structure functions  taken to be $F_{2N}^{Weak}(x,Q^2)$ and $F_{3N}^{Weak}(x,Q^2)$. 
 
 The study of the nucleon structure functions gives important information about the structure of the nucleon and provides opportunity to test the predictions of the perturbative Quantum Chromodynamics(QCD). Depending upon the kinematic region of the centre of mass energy ($W$) and the four momentum transfer square($Q^2$), it can also provide some important information about the non-perturbative QCD. In case of the electromagnetic(EM) DIS reactions induced by electrons and muons there is considerable experimental data on the EM nucleon structure functions viz. $F_{1N}^{EM}(x,Q^2)$ and  $F_{2N}^{EM}(x,Q^2)$ and the nuclear structure functions $F_{1A}^{EM}(x,Q^2)$ and  $F_{2A}^{EM}(x,Q^2)$ enabling us to study the nuclear medium effects on the electromagnetic nucleon structure functions by making a comparative study. This is not so in the case of weak nucleon structure function $F_{1N}^{Weak}(x,Q^2)$, $F_{2N}^{Weak}(x,Q^2)$ and  $F_{3N}^{Weak}(x,Q^2)$, 
where there is almost no experimental data on free nucleons. The weak nucleon structure functions have to be extracted from the (anti)neutrino DIS experiment on heavy nuclear targets like freon, freon-propane, etc. 
MINERvA at Fermilab~\cite{Mousseau:2016snl} has performed experiment with $\nu_\mu$ and $\bar\nu_\mu$ beams in the wide range of $x$ and $Q^2$ using several nuclear targets like carbon, iron, lead, etc. and the aim is to understand nuclear medium effects by doing EMC kind of measurements. Theoretically, a better understanding of nuclear medium effects on the weak structure functions is also required.

Generally there are two approaches in order to understand the medium effects in the weak nuclear structure functions, one is phenomenological
and the other is theoretical. In the phenomenological analysis, the general approach is that nuclear parton distribution 
functions(PDFs ) are obtained using the charged lepton-nucleus scattering data and analyzing the ratio of the structure 
 functions e.g. $\frac{F_{2A}^{EM}}{F_{2A^\prime}^{EM}}$,  $\frac{F_{2A}^{EM}}{F_{2D}^{EM}}$, where $A, A^\prime$ represent any two nuclei and $D$ stands for the deuteron,
 nuclear correction factor is determined. The same correction factor is then used for the weak structure functions 
 $F_{1A}^{Weak}(x,Q^2)$ and $F_{2A}^{Weak}(x,Q^2)$.
% The nuclear correction factor is then multiplied with free nucleon PDFs to get nuclear PDFs. 
% Therefore, for $F_{i\;\;A}^{\text{Weak}}(x,Q^2 )$ (i=1,2), same nuclear effect as that for $F_{i\;\;A}^{\text{EM}}(x,Q^2 )$ is assumed.
 For $F_{3A}^{\text{Weak}}(x,Q^2 )$, the information is inferred from (anti)neutrino scattering data~\cite{Berge:1989hr,Oltman:1992pq,Tzanov:2005kr}. The other phenomenological approach 
 is to directly extract the nuclear PDFs by analyzing the experimental data i.e. without using nucleon PDFs or nuclear correction factor. This approach has been
 recently used by nCTEQ~\cite{Kovarik:2015cma} group in getting $F_{2A}^{EM}(x,Q^2)$, $F_{2A}^{Weak}(x,Q^2)$ 
 and $F_{3A}^{Weak}(x,Q^2)$ nuclear structure functions by analyzing together the 
   charged lepton-$A$ DIS data and Drell-Yan $p$-$A$ data sets, and separately analyzing $\nu(\bar \nu)-A$ DIS data sets. 
   Their observation is that the nuclear medium effects on $F_{2A}^{EM}(x,Q^2)$ in
 electromagnetic interaction are different from $F_{2A}^{Weak}(x,Q^2)$ in the weak interaction specially at low $x$.

On the other hand, theoretically, there have been very few calculations to study nuclear medium effects in the weak structure functions. 
One is by us (Aligarh-Valencia collaboration~\cite{Haider:2011qs,Haider:2012nf,Haider:2016zrk,Zaidi:2019mfd,Zaidi:2019asc,Ansari:2020xne,SajjadAthar:2009cr}), and 
 the other is by Kulagin and Petti~\cite{Kulagin:2007ju,Kulagin:2004ie}. The present review is based on our theoretical works performed in the last 
 several years~\cite{Haider:2011qs,Haider:2012nf,Haider:2016zrk,Zaidi:2019mfd,Zaidi:2019asc,Ansari:2020xne,SajjadAthar:2009cr},
 on the electromagnetic and weak interaction induced DIS processes on the free nucleon and nuclear target in the wide region of $x$ and $Q^2$. 
 Recently we have extended our study for $\nu_\tau/\bar\nu_\tau$ scattering on the free nucleon and obtained the structure functions 
 as well as the differential and total scattering cross sections by including various perturbative and non-perturbative effects~\cite{Ansari:2020xne}. Work is in progress to understand nuclear medium effects in $\nu_\tau(\bar\nu_\tau)$-nucleus interactions and will be reported elsewhere.
% In this work we shall also present the results of our study on the nuclear structure functions by taking medium effects into account.
 
 In the present paper, we review the work on the nuclear medium effects in the structure functions using a microscopic approach based on field theoretical formalism~\cite{Marco:1995vb}.
 A relativistic nucleon spectral function has been used to describe the energy and momentum distribution of the nucleons in
 nuclei ~\cite{FernandezdeCordoba:1991wf}. This is obtained by using the Lehmann's representation for the relativistic nucleon propagator and 
 nuclear many body theory is used to calculate it for an interacting Fermi sea in nuclear matter. A local density approximation is then applied to 
 translate these results to finite nuclei. Furthermore, we include the contributions of meson clouds and include the pion and rho contributions in a many 
 body field theoretical approach. For the shadowing and anti-shadowing corrections which have been found to be effective in the low region of $x$($x \le 0.2$), we follow the works of
  Kulagin and Petti~\cite{Kulagin:2007ju,Kulagin:2004ie}. In the present work the numerical results for various structure functions and the cross sections have been presented and compared with experiments. Predictions have been made for $^{40}Ar$, relevant for the upcoming DUNE experiment\cite{Abi:2018alz} at the Fermilab.
 
The plan of the paper is the following. In section-2, we describe, in brief the formalism for calculating $\nu_l({\bar\nu}_l)$-nucleon scattering cross section and in section-3 we discuss the nuclear medium effects in the evaluation of structure functions and differential scattering cross section for $\nu_l({\bar\nu}_l)$-nucleus scattering. In section-4, we present and discuss the results and section-5, summarizes our study.
\section{Formalism}
 \subsection{(Anti)neutrino-nucleon cross section and the structure functions}\label{sec_formalism_N}
 The basic reaction for the (anti)neutrino induced charged current deep inelastic scattering process on a free nucleon target is given by
\begin{eqnarray}\label{reaction}
 \nu_l(k) / \bar\nu_l(k) + N(p) \rightarrow l^-(k') / l^+(k') + X(p');\;\;\;l=e/\mu/\tau,\;\;N=n,p,
\end{eqnarray}
where $k$ and $k'$ are the four momenta of incoming and outgoing lepton, $p$ and $p'$ are the four momenta of the target nucleon and the jet of
hadrons produced in the final state, respectively. 
This process is mediated by the $W$-boson ($W^\pm$) and the invariant matrix element corresponding to the reaction given in Eq.\ref{reaction}, is written as
\begin{equation}\label{matrix}
 -i{\cal M}=\frac{iG_F}{\sqrt{2}}\;l_\mu \;\left(\frac{M_W^2}{q^2-M_W^2} \right)\;\langle X|J^\mu|N\rangle\;,
\end{equation}
where $G_F$ is the Fermi coupling constant, $M_W$ is the mass of $W$ boson, and $q^2=(k-k')^2$ is the four momentum transfer square. $l_\mu$ is the leptonic 
current and $\langle X|J^\mu|N\rangle$ is the 
hadronic current for the neutrino induced reaction. 

\begin{figure}[h]
\begin{center}
\includegraphics[height=4.5 cm, width=8.5 cm]{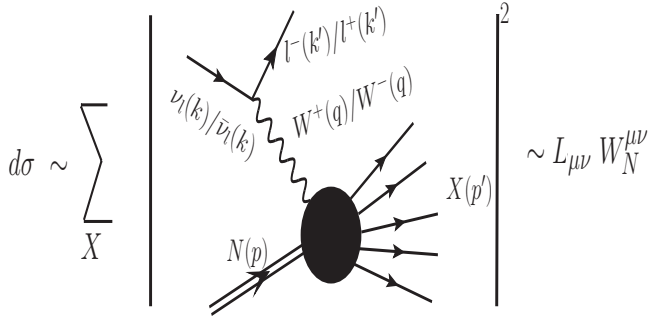}
\end{center}
\caption{ $\nu_l({\bar\nu}_l) - N$ inclusive scattering, where the 
 summation sign represents the sum 
over all the hadronic states such that the cross section($d\sigma$) for the deep inelastic scattering  
$\propto L_{\mu \nu} W_{N}^{\mu \nu}$.}
\label{fg:fig1}
\end{figure}

The general expression of the double differential scattering cross section (DCX) corresponding to the reaction given in Eq.~\ref{reaction} (depicted in Fig.~\ref{fg:fig1}) in the laboratory frame is expressed as:
\begin{equation}
\label{eq:w1w2w3}
\frac{ d^2\sigma  }{ dx dy } =  \frac{y M_N}{\pi }~\frac{E_\nu}{E_l}~\frac{|{\bf k^\prime}|}{|{ \bf k}|}\;  \frac{G_F^2}{2}~\left(\frac{M_W^2}{Q^2+M_W^2}\right)^2 ~L_{\mu\nu} ~W^{\mu\nu}_N, 
\end{equation}
where $x$ and $y$ are the scaling variables which lie in the following ranges:
\begin{equation}
\frac{m_l^2}{2M_N (E_\nu - m_l)} \le x \le 1;~~~a-b \le y\le a+b,
\end{equation}
with
\begin{eqnarray}
 a&=&\frac{1-m_l^2\Big(\frac{1}{2M_NE_\nu x}+\frac{1}{2E_\nu^2} \Big)}{2\Big(1+\frac{M_N x}{2E_\nu}\Big)},\;\;
 b=\frac{\sqrt{\left(1-\frac{m_l^2}{2 M_N E_\nu x}\right)^2-\frac{m_l^2}{E_\nu^2}}}{2\Big(1+\frac{M_N x}{2E_\nu}\Big)},
\end{eqnarray}
 where $m_l$ is the charged lepton mass. The leptonic tensor $L_{\mu\nu} $ is given by  
\begin{eqnarray}\label{lep_weak}
L_{\mu \nu} &=&8(k_{\mu}k'_{\nu}+k_{\nu}k'_{\mu}
-k.k^\prime g_{\mu \nu}  \pm i \epsilon_{\mu \nu \rho \sigma} k^{\rho} 
k'^{\sigma})\,,
\end{eqnarray}
 with +ve sign
for antineutrino and -ve sign for neutrino in the antisymmetric term. 

The hadronic tensor $W_{N}^{\mu \nu}$ is written in terms of the weak nucleon structure functions $W_{iN} (\nu,Q^2)~(i=1-6)$ as
\begin{eqnarray}\label{ch2:had_ten_N}
W_{N}^{\mu \nu} &=&
\left( \frac{q^{\mu} q^{\nu}}{q^2} - g^{\mu \nu} \right) \;
W_{1N} (\nu, Q^2)
+ \frac{W_{2N} (\nu, Q^2)}{M_N^2}\left( p^{\mu} - \frac{p . q}{q^2} \; q^{\mu} \right)
 \nonumber\\
&\times&\left( p^{\nu} - \frac{p . q}{q^2} \; q^{\nu} \right)-\frac{i}{2M_N^2} \epsilon^{\mu \nu \rho \sigma} p_{ \rho} q_{\sigma}
W_{3N} (\nu, Q^2) + \frac{W_{4N} (\nu, Q^2)}{M_N^2} q^{\mu} q^{\nu}\nonumber\\
&&+\frac{W_{5N} (\nu, Q^2)}{M_N^2} (p^{\mu} q^{\nu} + q^{\mu} p^{\nu})
+ \frac{i}{M_N^2} (p^{\mu} q^{\nu} - q^{\mu} p^{\nu})
W_{6N} (\nu, Q^2)\,.
\end{eqnarray}
% where $W_{iN} (\nu, Q^2);~(i=1-6)$ are the nucleon structure functions and $\nu(=k^0-k^{'0})$ is the energy transfer. 
The contribution of the term with $W_{6N} (\nu, Q^2)$ vanishes when contracted with the leptonic tensor. In the limit of
high $Q^2$ and $\nu$, the structure functions $W_{iN}  (\nu,Q^2);~(i=1-5)$ are generally expressed in terms of the dimensionless nucleon structure functions $F_{iN}  (x),\;\;i=1 - 5$. However, as we move towards the region of low and moderate $Q^2$, the dimensionless nucleon structure functions show $x$ as well as $Q^2$ dependence. Hence, $F_{iN}  (x,Q^2),\;\;i=1 - 5$ are defined as
\begin{small}
 \begin{eqnarray}\label{ch2:relation}
 F_{1N}(x,Q^2) &=& W_{1N}(\nu,Q^2),\;\;F_{2N}(x,Q^2) = \frac{Q^2}{2xM_N^2}W_{2N}(\nu,Q^2),\;\;
 F_{3N}(x,Q^2) = \frac{Q^2}{xM_N^2}W_{3N}(\nu,Q^2),\nonumber\\
 F_{4N}(x,Q^2) &=& \frac{Q^2}{2M_N^2}W_{4N}(\nu,Q^2),\;\; 
 F_{5N}(x,Q^2) = \frac{Q^2}{2xM_N^2}W_{5N}(\nu,Q^2).\;\;\;
%  \{F_{1N},F_{2N},F_{3N},F_{4N},F_{5N}\}=\{W_{1N},\frac{Q^2}{2xM_N^2}W_{2N},\frac{Q^2}{xM_N^2}W_{3N},\frac{Q^2}{2M_N^2}W_{4N},\frac{Q^2}{2xM_N^2}W_{5N}\}.\nonumber
\end{eqnarray}
\end{small}
 The expression for the differential scattering cross section for the $\nu_l/{\bar\nu}_l - N$ scattering 
given in Eq.~\ref{eq:w1w2w3} is written by using Eqs.~\ref{lep_weak} and \ref{ch2:relation} as:
\begin{small}
\begin{eqnarray}
 \frac{d^2\sigma}{dxdy}&=&\frac{G_F^2M_NE_\nu}{\pi(1+\frac{Q^2}{M_W^2})^2}
 \Big\{\Big[y^2x+\frac{m_l^2 y}{2E_\nu M_N}\Big]F_{1N}(x,Q^2)+
 \Big[\Big(1-\frac{m_l^2}{4E_\nu^2}\Big)-\Big(1+\frac{M_Nx}{2E_\nu}\Big)y\Big]F_{2N}(x,Q^2)\nonumber\\
 &\pm& \Big[xy\Big(1-\frac{y}{2}\Big)-
 \frac{m_l^2 y}{4E_\nu M_N}\Big]F_{3N}(x,Q^2)
 +\frac{m_l^2(m_l^2+Q^2)}{4E_\nu^2M_N^2 x}F_{4N}(x,Q^2)-\frac{m_l^2}{E_\nu M_N}F_{5N}(x,Q^2)\Big\}.\;\;\;~~~~~
\end{eqnarray}
\end{small}
% 
% are written in terms of PDFs by using the well known Callan-Gross and Albright-Jarlskog relations, respectively:

In general, the dimensionless nucleon structure functions are derived in the quark-parton model assuming Bjorken scaling in which they  scale and are functions of only one variable $x$. In this limit, these structure functions obey Callan-Gross~\cite{Callan:1969uq} and Albright-Jarlskog~\cite{Albright:1974ts} relations 
\begin{eqnarray}
 F_{1}(x)&=&\frac{F_2(x)}{2 x}\;;\;\;
 F_{5}(x)=\frac{F_2(x)}{2 x}, \nonumber
\end{eqnarray}
and at the leading order, the structure functions are written in terms of the parton distribution functions $q_i(x)$ and $\bar q_i(x)$ as:
\begin{eqnarray}\label{parton_wk}
F_{2}(x)  &=& \sum_{i} x [q_i(x) +\bar q_i(x)] \;;\;
x F_3(x) =  \sum_i x [q_i(x) -\bar q_i(x)]\;;\;\;
F_4(x)=0.
\end{eqnarray} 
 For example, in the case of $\nu(\bar{\nu})$-proton scattering 
 above the charm production threshold, $F_{2,3}(x)$ are given by:
\begin{eqnarray}\label{eq:pdf1}
F_{2p}^{\nu }(x)& = & 2 x [d(x) + s(x) + \bar{u}(x) +\bar{c}(x)]\;;\;\; F_{2p}^{\bar\nu}(x)=  2 x [u(x) + c (x)+ \bar{d}(x) +\bar{s}(x)];\nonumber\\
x F_{3p}^{\nu}(x)& = & 2 x [d(x) + s (x)- \bar{u}(x) -\bar{c}(x)]\;;\;\;
x F_{3p}^{\bar\nu}(x)= 2 x [u(x) + c(x) - \bar{d}(x) -\bar{s}(x)]\;\;\;\;\;\;\;~~
\end{eqnarray}
and for the $\nu(\bar{\nu})$-neutron scattering $F_{2,3}(x)$ are given by
\begin{eqnarray}\label{eq:pdf2}
F_{2n}^{\nu }(x) & = & 2 x [u(x) + s(x) + \bar{d}(x) +\bar{c}(x)]\;;\;\;F_{2n}^{\bar\nu}(x)  =  2 x [d (x)+ c(x) + \bar{u}(x) +\bar{s}(x)];\nonumber\\
x F_{3n}^{\nu}(x) & = & 2 x [u(x) + s (x)- \bar{d}(x) -\bar{c}(x)]\;;\;\;
x F_{3n}^{\bar\nu}(x)  =  2 x [d(x) + c(x) - \bar{u}(x) -\bar{s}(x)].\;\;\;\;\;\;\;
\end{eqnarray}
For an isoscalar nucleon target, we use
\begin{equation}
 F_{iN}=\frac{F_{ip}+F_{in}}{2}.~~~(i=1-5)
\end{equation}

In the present formalism we have performed numerical calculations in the three($u$, $d$ and $s$) as well as four($u$, $d$, $s$ and $c$) flavor schemes by taking 
$u$, $d$ and $s$ to be massless and the charm quark to be massless as well as massive. 
In the region of low and moderate $Q^2$, the perturbative 
 and nonperturbative QCD corrections such as $Q^2$ evolution of parton distribution functions from leading order to 
 higher order terms 
 (next-to-leading order (NLO), next-next-to-leading order (NNLO), ...), the effects of target mass correction due to 
 the massive quarks production (e.g. charm, bottom, top) and higher twist (twist-4, twist-6, ...) because of the multiparton correlations, become important. 
 These nonperturbative effects are specifically important in the kinematical region of high $x$ and low $Q^2$. 
  The $Q^2$ evolution of
structure functions is determined by the DGLAP evolution equation~\cite{Dokshitzer:1977sg,Gribov:1972ri,Altarelli:1977zs,Lipatov:1974qm}. 
The parton distribution functions for the nucleon have been determined by 
various groups and we have taken the parameterization of MMHT~\cite{Harland-Lang:2014zoa} in our numerical calculations
 up to NNLO following Ref.~\cite{Vermaseren:2005qc,Moch:2004xu,Moch:2008fj}.
The nonperturbative higher twist
 effect is incorporated by using the renormalon approach~\cite{Dasgupta:1996hh} and the target mass correction is included following the works 
 of Schienbein et al.~\cite{Schienbein:2007gr}. The incorporation of the contribution from gluon emission induces the $Q^2$ dependence of the nucleon structure functions.
  The details of the discussion are given in Ref.~\cite{Zaidi:2019asc}.
  \subsection{(Anti)neutrino-nucleus cross section and structure functions}\label{sec_formalism_A}
\begin{figure}
\begin{center}
\includegraphics[height=5 cm, width=7 cm]{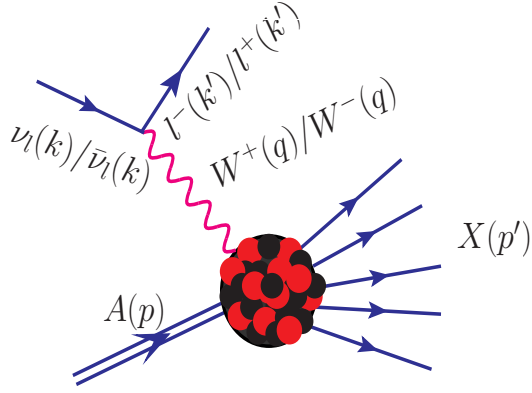}
\end{center}
\caption{ Feynman diagrams for the $\nu_l/\bar\nu_l;~(l=e, \mu,\tau)$ induced DIS process off nuclear target ($A$).}
\label{dis_nuc}
\end{figure}

% \section{Formalism: $\nu_l-A$ DIS process}\label{sec_formalism_A}
  The differential scattering cross section for the charged current inclusive $\nu_l/\bar\nu_l$-nucleus deep inelastic scattering process (depicted in Fig.~\ref{dis_nuc}):
\begin{equation}\label{DISrxn}
 \nu_l / \bar\nu_l(k) + A(p_A) \rightarrow l^-/l^+(k') + X(p'_A)
\end{equation}
is expressed in terms of the leptonic tensor $L_{\mu\nu}$ and 
  the nuclear hadronic tensor $W^{\mu\nu}_A$ as:
\begin{eqnarray}\label{xecA}
{ d^2\sigma_A \over dx dy }&=& \left({G_F^2 y M_N E_\nu  \over 2\pi E_{l}}\right) {\left(M_W^2\over M_W^2+Q^2 \right)^2} \; {|{\bf k^\prime}| \over|{ \bf k}|}\;
L_{\mu\nu}\; W^{\mu\nu}_A,
\end{eqnarray}
where the physical quantities have their usual meanings. The expression of $L_{\mu\nu}$ is given by Eq.\ref{lep_weak}. 
The nuclear hadronic tensor $W^{\mu\nu}_A$ is written in terms of the weak nuclear structure functions $W_{iA}(\nu_A,Q^2)$ ($i=1-6$) as:
\begin{eqnarray}
 \label{nuc_had_weak}
W_{A}^{\mu \nu} &=&
\left( \frac{q^{\mu} q^{\nu}}{q^2} - g^{\mu \nu} \right) \;
W_{1A} (\nu_A, Q^2)
+ \frac{W_{2A} (\nu_A, Q^2)}{M_A^2}\left( p^{\mu}_A - \frac{p_A . q}{q^2} \; q^{\mu} \right)
 \nonumber\\
&\times&\left( p^{\nu}_A - \frac{p_A . q}{q^2} \; q^{\nu} \right)\pm\frac{i}{2M_A^2} \epsilon^{\mu \nu \rho \sigma} p_{A \rho} q_{\sigma}
W_{3A} (\nu_A, Q^2) + \frac{W_{4A} (\nu_A, Q^2)}{M_A^2} q^{\mu} q^{\nu}\nonumber\\
&&+\frac{W_{5A} (\nu_A, Q^2)}{M_A^2} (p^{\mu}_A q^{\nu} + q^{\mu} p^{\nu}_A)
+ \frac{i}{M_A^2} (p^{\mu}_A q^{\nu} - q^{\mu} p^{\nu}_A)
W_{6A} (\nu_A, Q^2)\,,
\end{eqnarray}
where $M_A$ is the mass of the nuclear target. 
 The nuclear structure functions $W_{iA}(\nu_A,Q^2)~(i=1-5)$ are written in terms of the dimensionless nuclear structure functions $F_{iA}(x_A,Q^2)~(i=1-5)$ as: 
 \begin{small}
\begin{eqnarray}\label{relation1}
% \left.
% \begin{array}{l}
 F_{1A}(x_A,Q^2) &=& W_{1A}(\nu_A,Q^2)\;;\;\; F_{2A}(x_A,Q^2) = \frac{Q^2}{2xM_A^2}W_{2A}(\nu_A,Q^2)\;;\;\; F_{3A}(x_A,Q^2) = \frac{Q^2}{xM_A^2}W_{3A}(\nu_A,Q^2);\nonumber\\
 F_{4A}(x_A,Q^2) &=& \frac{Q^2}{2M_A^2}W_{4A}(\nu_A,Q^2) \;;\;\;
 F_{5A}(x_A,Q^2) = \frac{Q^2}{2xM_A^2}W_{5A}(\nu_A,Q^2), 
%  \end{array}
%  \right\}
\end{eqnarray}
\end{small}
where $\nu_A$(=$\frac{p_{_A}\cdot q}{M_{_A}}(=q^{0})$ ) is the energy transferred to the target in the Lab frame and $x_A(=x/A)$ is the Bjorken scaling variable given by:
\begin{eqnarray}
x_A&=&\frac{Q^2}{2 p_A \cdot q}=\frac{Q^2}{2 p_{A}^0  q^0 } = \frac{Q^2}{2 A~M_N q^0}=\frac{x}{A}.
\end{eqnarray} 
% $W_{6A}(\nu_A,Q^2)$ does not contribute to the cross section as it vanishes when contracted with the leptonic tensor $L_{\mu \nu}$.

The expression for the differential cross section for the $\nu_l/{\bar\nu}_l - A$ scattering can be obtained using Eqs.~\ref{lep_weak},  \ref{nuc_had_weak} and \ref{relation1} in Eq.~\ref{xecA} as
\begin{small}
\begin{eqnarray}\label{xsecsf}
 \frac{d^2\sigma_A}{dxdy}&=&\frac{G_F^2M_NE_\nu}{\pi(1+\frac{Q^2}{M_W^2})^2}
 \Big\{\Big[y^2x+\frac{m_l^2 y}{2E_\nu M_N}\Big]F_{1A}(x,Q^2)+
 \Big[\Big(1-\frac{m_l^2}{4E_\nu^2}\Big)-\Big(1+\frac{M_Nx}{2E_\nu}\Big)y\Big]F_{2A}(x,Q^2)\nonumber\\
 &\pm& \Big[xy\Big(1-\frac{y}{2}\Big)-
 \frac{m_l^2 y}{4E_\nu M_N}\Big]F_{3A}(x,Q^2)
 +\frac{m_l^2(m_l^2+Q^2)}{4E_\nu^2M_N^2 x}F_{4A}(x,Q^2)-\frac{m_l^2}{E_\nu M_N}F_{5A}(x,Q^2)\Big\}.~~~\;\;\;
\end{eqnarray}
\end{small}
 
For $\nu_e/\bar{\nu}_e$ and $\nu_\mu/\bar{\nu}_\mu$ charged current interactions, in the limit $m_l\to 0$, only the first three terms of Eq.~\ref{xsecsf}, i.e. the terms with $F_{1A}(x,Q^2)$, $F_{2A}(x,Q^2)$ and $F_{3A}(x,Q^2)$ structure functions would contribute.
We consider the scattering process in the laboratory frame, where target nucleus is at rest i.e. $p_A=(p_A^0,~{\bf p_A}= 0)$ and the momentum of the nucleon in the nucleus (${\bf p_N}$) is non-zero and the motion of such nucleons corresponds to the 
Fermi motion.

 If the momentum transfer is along the $Z$-axis then $q^\mu=(q^0,0,0,q^z)$ and the Bjorken variable $x_N$ is written as:
 \begin{equation}
 x_N = \frac{Q^2}{2 p_N \cdot q} = \frac{Q^2}{2 (p_N^0 q^0 - p_N^z q^z)}.
\end{equation}
%The bound nucleons interact with each other through the strong force hence various nuclear medium effects come into the picture. Depending upon the value of the Bjorken variable $x$ the various nuclear medium effects have different contribution.
 %Fermi motion, binding energy, nucleon correlations, mesonic contributions, etc. .
  The nuclear medium effects such as Fermi motion, binding, nucleon correlations incorporated through nucleon spectral function, meson cloud contribution and shadowing effect are discussed in the following subsections \ref{spec}, \ref{mes} and \ref{sec_shad}.
\subsubsection{Effect of Fermi motion, binding and nucleon correlation}\label{spec}
% In order to calculate the cross section for the neutrino scattering off a bound nucleon in the nuclear medium, we begin with a neutrino flux hitting  the target  nucleons over a given period of time. Since neutrinos are weakly interacting particles so the majority of the neutrinos will  pass through the target without interacting while a few neutrinos will
%  interact with the target nucleons yielding final state leptons and
%  hadrons.  Here we put forward the concept of ``neutrino self energy". The real part of ``neutrino self energy" modifies the lepton mass and imaginary part  gives information about the total number of neutrinos interactions that yield the final state leptons and hadrons.
%  The basic ingredients of the model are given in Appendix~\ref{app:self}-\ref{app:lda}.

 We start by writing the scattering cross section ($d\sigma_A$) for small elemental volume ($dV$) inside the nucleus in terms of the probability of neutrino interaction with a bound nucleon per unit time ($\Gamma$). Probability times the differential of area ($dS$) defines the cross section~\cite{Marco:1995vb}, i.e.
%   The cross section for an element of volume $dV$ in the nucleus is defined as~\cite{Marco:1995vb}:
\begin{equation}\label{defxsec}
d\sigma_A=\Gamma dt dS = \Gamma \frac {E_\nu}{\mid   {\bf k}  \mid}d^3 r,~~~~~~~~~\Big[\because~~dtdS= \frac{dV}{v}=\frac {E_\nu}{\mid {\bf k}\mid }d^3 r \Big]
% =\frac{-2m_l}{E_l({\bf k})} Im \Sigma (k)\frac{E_l({\bf k})}{\mid {\bf k} \mid}dV,
\end{equation}
where $v$ is the velocity of the incoming neutrino. $\Gamma$ is related to the imaginary part of the $\nu_l$ self energy ($\Sigma(k)$) as~\cite{Marco:1995vb}:
\begin{equation}\label{eqr}
-\frac{\Gamma}{2} = \frac{m_\nu}{E_\nu({\bf k})}\; Im \Sigma(k).
\end{equation}
It may be pointed out that the neutrino self energy $\Sigma (k)$ has two parts, the real part of ``neutrino self energy" modifies the lepton mass and imaginary part i.e. $Im \Sigma (k)$ gives information about the total number of neutrinos interactions that yield the final state leptons and hadrons. 

From Eq.\ref{defxsec} and Eq.\ref{eqr}, we get
\begin{equation}\label{eqq}
 d\sigma_A= -2\frac{m_\nu}{\mid {\bf k} \mid} Im \Sigma (k) d^3 r.
\end{equation}

\begin{figure}
\begin{center}
 \includegraphics[height=4.0 cm, width=11 cm]{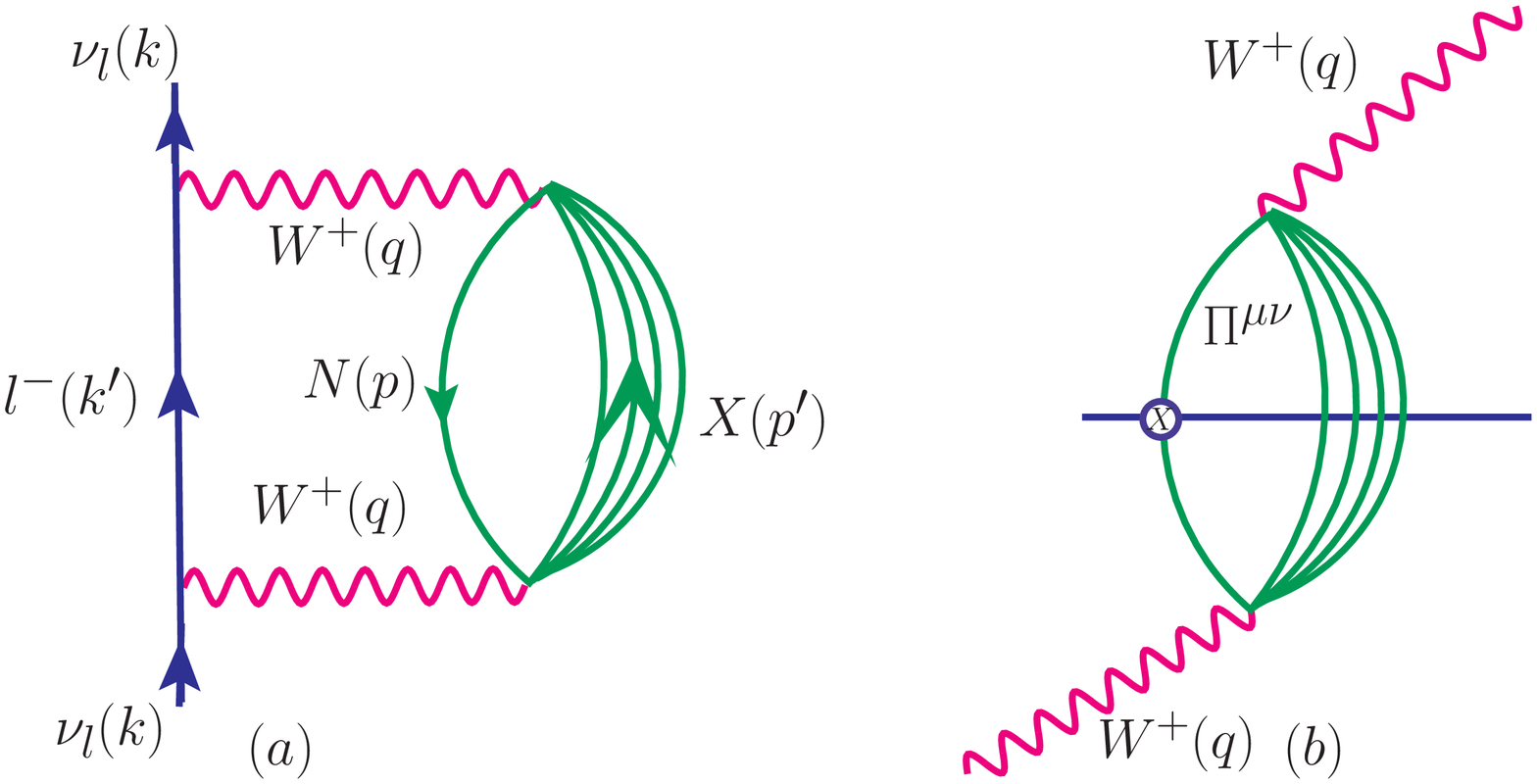}
 \end{center}
 \caption{Diagrammatic representation of {\bf (a)} the neutrino ($\nu_l$) self-energy and {\bf (b)} the intermediate vector boson ($W$) self-energy.}
 \label{wself_energy}
\end{figure}
 $\Sigma (k)$ is evaluated corresponding to the diagram shown in Fig.\ref{wself_energy} (left panel) using the Feynman rules
and on applying the Cutkowsky rules
 the imaginary part of neutrino self energy is obtained as~\cite{Haider:2016zrk}:
\begin{equation}\label{nu_imslf1}
Im \Sigma(k)=\frac{ G_F}{\sqrt{2}} {4 \over m_\nu} \int \frac{d^3 k^\prime}{(2 \pi)^4} {\pi \over E({\bf k^\prime})} \theta(q^0) \left(\frac{M_W}{Q^2+M_W^2}\right)^2\;Im[L_{\mu\nu}^{WI} \Pi^{\mu\nu}(q)],
\end{equation}
where $\Pi^{\mu\nu}(q)$ is the $W$-boson self-energy (as shown in Fig.~\ref{wself_energy}(b)), which is generally written in terms of the nucleon propagator ($G_l$)  and meson propagator ($D_j$) corresponding to  Fig.~\ref{wself_energy}(b), as:
\begin{eqnarray}\label{wboson}
 \Pi^{\mu \nu} (q)&=& \left(\frac{G_F M_W^2}{\sqrt{2}}\right) \times \int \frac{d^4 p}{(2 \pi)^4} G (p) 
\sum_X \; \sum_{s_p, s_l} \prod_{i = 1}^{N} \int \frac{d^4 p'_i}{(2 \pi)^4} \; \prod_{_l} G_l (p'_l)\prod_{_j} \; D_j (p'_j)\; \nonumber \\  
&&  <X | J^{\mu} | N >  <X | J^{\nu} | N >^* (2 \pi)^4 ~
\delta^4 \Big(k + p - k^\prime - \sum^N_{i = 1} p'_i\Big),\;\;\;
\end{eqnarray}
where $s_p$ and $s_l$ are the spin of the initial state nucleon and the final state fermions, the indices $l$ and $j$ are respectively, for the fermions and bosons in the final hadronic state, $<X | J^{\mu} | N>$ represents the hadronic current and $\delta^4 (k + p - k^\prime - \sum^N_{i = 1} p'_i)$ ensures the conservation
of four momentum. $G(p)$ gives the information about the propagation of the nucleon from the initial state to the final state or vice versa. 

 To obtain the relativistic nucleon propagator $G(p^0,{\bf p})$ in a nuclear medium we start with 
 the relativistic free nucleon Dirac propagator $G^{0}(p^{0},{\bf p})$ which is written in terms of the Dirac spinors $u({\bf p})$ and $v({\bf p})$.
 This includes the contribution from positive and negative energy components of the nucleon and the negative energy 
 contribution is suppressed while the positive energy contribution survives~\cite{Marco:1995vb,FernandezdeCordoba:1991wf}.
%  Therefore, the free nucleon propagator may be expressed as
% \begin{eqnarray}
%  G^{0}(p^{0},{{\bf p}}) = \frac{1}{\not p - M_N+i \epsilon}=\frac{\not p+M_N}{(p^2-M_N^2+i\epsilon)}. 
% \end{eqnarray}
 Considering only the positive energy part of the nucleon propagator $G^{0}(p^{0},{{\bf p}})$, we write
 \begin{equation}
 G^{0}(p^{0},{{\bf p}})= {\not p +M_N \over p^2-M_N^2 +i\epsilon} + 2\ i \pi \theta(p^0) \delta(p^2 -M_N^2) n({\bf p}) (\not p +M_N).
\end{equation}
 In the nuclear medium considered as an interacting Fermi sea, $G(p^0,{\bf p})$ is written
in terms of the 
nucleon self energy $\Sigma^N(p^0,\bf{p})$, which contains all the information on single nucleon. Then in nuclear medium the interaction is taken into account through Dyson series expansion, which is in principle an infinite series in perturbation theory. 
This perturbative expansion is summed in a ladder approximation as~\cite{Marco:1995vb}:
% We add 
% this perturbative expansion in a ladder approximation (Fig.\ref{n_self}) as: 
% \begin{eqnarray}\label{gpseries}
%  G(p) &=& G^0(p)~+~G^0(p)\Sigma^N(p)G^0(p)~+~G^0(p)\Sigma^N(p)G^0(p)\Sigma^N(p)G^0(p)~+~.......,\nonumber
% \end{eqnarray}
% which after simplification modifies to
\begin{eqnarray}\label{gp1}
G(p)&=&
% \frac{M_N}{E({\bf p})}\frac{\sum_{r}u_{r}({\bf p})\bar u_{r}({\bf p})}{p^{0}-E({\bf p})}+\frac{M_N}{E({\bf p})}\frac{\sum_{r}u_{r}({\bf p})\bar
% u_{r}({\bf p})}{p^{0}-E({\bf p})}\Sigma^N(p^{0},{\bf p})
%  \frac{M_N}{E({\bf p})} \frac{\sum_{s}u_{s}({\bf p})\bar u_{s}({\bf p})}{p^{0}-E(P)}+..... \nonumber \\
 \frac{M_N}{E({\bf p})}\frac{\sum_{r} u_{r}({\bf p})\bar u_{r}({\bf p})}{p^{0}-E({\bf p})-\sum_{r}\bar u_{r}({\bf p})\Sigma^N (p^{0},{\bf p})u_{r}({\bf p})
\frac{M_N}{{E({\bf p})}}}.\;\;\;
\end{eqnarray}
One may notice from the expression for the nucleon propagator $G(p)$ given in Eq.\ref{gp1} that it contains nucleon self energy $\Sigma^N(p_0,{\bf p})$.   
 The nucleon self-energy is written using the techniques of the standard Many-Body Theory~\cite{FernandezdeCordoba:1991wf}. 
 The inputs required for the NN interaction are incorporated by relating them to the experimental elastic NN cross section. 
  Furthermore, RPA-correlation effect is taken into account using the spin-isospin effective 
 interaction as the dominating part of the particle-hole (ph) interaction. Using the modified expression for the nucleon self energy, the imaginary part of 
 it is obtained~\cite{Haider:2015vea}.
%  The Fock term, which does not have imaginary part, neither contributes to real or imaginary part of nucleon self energy.
% Its contribution to the nucleon self energy is explicitly calculated and added to $Re \Sigma^N(p^{0},{\bf p})$~\cite{FernandezdeCordoba:1991wf}.
%  However this model  misses some contributions from similar Hartree type of terms which are independent of the momentum of nucleon $p$. 
%  Nucleon self energy in this model is found to be in reasonable agreement with those obtained 
% in sophisticated  many body calculations and this model has been effectively used in the past for the study of nuclear medium effects
% in photons, pions and leptons induced processes ~\cite{Gil:1997bm, Gil:1997jg}.
These considerations lead to a dressed nucleon propagator in the nuclear matter which is given by~\cite{Marco:1995vb,FernandezdeCordoba:1991wf}:
%  \begin{small}
 \begin{eqnarray}\label{Gp}
G (p) =&& \frac{M_N}{E({\bf p})} 
\sum_r u_r ({\bf p}) \bar{u}_r({\bf p})
\left[\int^{\mu}_{- \infty} d \omega 
\frac{S_h (\omega, {\bf{p}})}{p^0 - \omega - i \eta}
+ \int^{\infty}_{\mu} d  \omega 
\frac{S_p (\omega, {\bf{p}})}{p^0 - \omega + i \eta}\right],\;\;~~~
\end{eqnarray}
where the expressions for the 
hole $S_h (\omega, {\bf{p}})$(for $p_0 \le \mu$) and the particle $S_p (\omega, {\bf{p}})$(for $p_0 \ge \mu$) 
spectral functions and the nucleon self energy $\Sigma^N(p^0,{\bf{p}})$ are taken from
 Ref.~\cite{FernandezdeCordoba:1991wf}. For an inclusive process, only the hole spectral function contributes. 
In the above expression, $\mu$ is the chemical potential given by $\mu=\frac{p_F^2}{2M}~+~Re\Sigma^N\left[\frac{p_F^2}{2M},p_F\right]$ and $\omega=p^0-M_N$ is the removal energy. $\eta$ is the infinitesimal quantity i.e. $\eta \to 0$. $p_F$ is the Fermi momentum of the nucleon in the nucleus. In the local Fermi gas model the Fermi momentum is a function $r$, the point at which the interaction in the nucleus takes place and is given by $p_{_{F}}(r)=\left[ \frac{3}{2}\pi^{2} \rho(r) \right]^{1/3}$, where $\rho(r)$ is the nucleon charge density inside the nucleus, the parameters of which are determined from electron scattering experiments~\cite{DeVries:1987}.
 The spectral function for an isoscalar nuclear target is normalized to the number of nucleons $(A)$ in the nucleus i.e.
\begin{eqnarray}
  4 \int d^3r\;\int \frac{d^3 p}{(2\pi)^3} \;\int_{-\infty}^{\mu}\;S_h(\omega,{\bf p},\rho(r))\;d\omega = A.\; 
%    2 \int d^3r\;\int \frac{d^3 p}{(2\pi)^3} \;\int_{-\infty}^{\mu_n}\;S_h^n(\omega,{\bf p},\rho_n(r))\;d\omega &=& N\;, \nonumber
 \end{eqnarray} 
 Some of the properties of the spectral function may be found in the appendix of Ref.~\cite{Haider:2015vea}.

Then using Eqs.~\ref{eqq} and \ref{nu_imslf1} the expression for the differential cross section is written as 
\begin{equation}\label{dsigma_3}
\frac {d\sigma_A}{dx dy}=-\frac{G_F^2\;M_N\;y}{2\pi}\;\frac{E_l}{E_\nu}\;\frac{|\bf{k^\prime}|}{|\bf {k}|}\left(\frac{M_W^2}{Q^2+M_W^2}\right)^2 L_{\mu\nu} \int  Im \Pi^{\mu\nu}(q) d^{3}r.
\end{equation}
Comparing Eq.\ref{dsigma_3}, with Eqs.\ref{xecA}, \ref{wboson} and \ref{Gp}, the nuclear hadronic tensor (for isospin symmetric nucleus) can be expressed in terms of 
the nucleon hadronic tensor and the hole spectral function and is given as~\cite{Haider:2015vea}
\begin{equation}\label{conv_WAa}
W^{\mu \nu}_{A} = 4 \int \, d^3 r \, \int \frac{d^3 p}{(2 \pi)^3} \, 
\frac{M_N}{E ({\bf p})} \, \int^{\mu}_{- \infty} d p^0 S_h (p^0, {\bf p}, \rho(r))
W^{\mu \nu}_{N} (p, q), \,
\end{equation}
a factor of 4 is because of the spin-isospin degrees of freedom of the nucleon. 

For a non-isoscalar nuclear target like $^{56}$Fe and $^{208}Pb$, the spectral functions for the proton ($Z$) and neutron ($N=A-Z$) numbers in a nuclear target which are the function of local Fermi momenta
 $p_{_{{F}_{p,n}}}(r)=\left[ 3\pi^{2} \rho_{p(n)}({r}) \right]^{1/3}$, are normalized separately such that
\begin{eqnarray}
\label{spec_normp}
  2 \int d^3r\;\int \frac{d^3 p}{(2\pi)^3} \;\int_{-\infty}^{\mu_p}\;S_h^p(\omega,{\bf p},\rho_p(r))\;d\omega &=& Z\;, \\
  \label{spec_normn}
    2 \int d^3r\;\int \frac{d^3 p}{(2\pi)^3} \;\int_{-\infty}^{\mu_n}\;S_h^n(\omega,{\bf p},\rho_n(r))\;d\omega &=& N\;, 
 \end{eqnarray}
where the factor of 2 is due to the two possible projections of nucleon spin, $\mu_p(\mu_n)$ is the chemical potential for the proton(neutron), and $S_h^p(\omega,{\bf p},\rho_p(r))$ and
$S_h^n(\omega,{\bf p},\rho_n(r))$ are the hole spectral functions for the proton and neutron, respectively.
 The proton and neutron densities $\rho_{p}(r)$ and $\rho_{n}(r)$ are related to the nuclear density $\rho(r)$ 
 as~\cite{Haider:2016zrk,Haider:2015vea}: 
\begin{eqnarray}
 \rho_{p}(r) &=& \frac{Z}{A}\;\rho(r)~;\hspace{5 mm}
 \rho_{n}(r) = \frac{(A-Z)}{A}\;\rho(r).\nonumber
\end{eqnarray}
Hence for a nonisoscalar nuclear target, the nuclear hadronic tensor is written as
\begin{equation}\label{conv_WAan}
W^{\mu \nu}_{A} = 2 \sum_{\tau=p,n} \int \, d^3 r \, \int \frac{d^3 p}{(2 \pi)^3} \, 
\frac{M_N}{E_N ({\bf p})} \, \int^{\mu_\tau}_{- \infty} d p^0 S_h^\tau (p^0, {\bf p}, \rho_\tau(r))\;
W^{\mu \nu}_{\tau} (p, q). \,
\end{equation}

To evaluate the weak dimensionless nuclear structure functions by using Eq.\ref{conv_WAa}, the appropriate components of nucleon ($W^{\mu\nu}_N$ in Eq.\ref{ch2:had_ten_N}) and nuclear ($W^{\mu\nu}_A$ in Eq.\ref{nuc_had_weak}) hadronic tensors along the $x$, $y$ and $z$ axis are chosen. For example, the expression of $F_{1A,N}(x_A,Q^2)$ is obtained by taking the $xx$ components, $F_{2A,N}(x_A,Q^2)$ by taking the $zz$ components, $F_{3A,N}(x_A,Q^2)$ by taking the $xy$ components, etc.~\cite{Zaidi:2019asc}, and 
 for an isoscalar nuclear target, the expressions for the three nuclear structure functions viz. $F_{iA,N}(x_A,Q^2)\;\;i=1-3$ i.e. for the massless leptons 
 as are obtained as: 
  \begin{eqnarray}\label{spect_funct}
F_{iA,N} (x_A, Q^2) &=& 4\int \, d^3 r \, \int \frac{d^3 p}{(2 \pi)^3} \, 
\frac{M_N}{E_N ({\bf p})} \, \int^{\mu}_{- \infty} d p^0~ S_h(p^0, {\bf p}, \rho(r))~
\times f_{iN}(x,Q^2),~~~\;\;\;
\end{eqnarray}
where
\begin{eqnarray}
 f_{1N}(x,Q^2)&=&AM_N\left[\frac{F_{1N} (x_N, Q^2)}{M_N} + \left(\frac{p^x}{M_N}\right)^2 \frac{F_{2N} (x_N, Q^2)}{\nu_N}\right],\\
f_{2N}(x,Q^2)&=& \left[\frac{Q^2}{(q^z)^2}\left( \frac{|{\bf p}|^2~-~(p^{z})^2}{2M_N^2}\right) +  \frac{(p^0~-~p^z~\gamma)^2}{M_N^2} \left(\frac{p^z~Q^2}{(p^0~-~p^z~\gamma) q^0 q^z}~+~1\right)^2\right]~\nonumber\\
&&~~~~\times\left(\frac{M_N}{p^0~-~p^z~\gamma}\right) \times F_{2 N} (x_N,Q^2),\\
% \end{eqnarray}
% \begin{eqnarray}
f_{3N}(x,Q^2)&=&A\frac{q^0}{q^z} \;\times\left({p^0 q^z - p^z q^0  \over p \cdot q} \right)F_{3N} (x_N,Q^2).
\end{eqnarray}
Similar expression is obtained for a non-isoscalar nuclear target~\cite{Zaidi:2019asc}.

 The nonperturbative effects of the target mass corrections and the higher twist~\cite{Dasgupta:1996hh} have been
  incorporated  in the free nucleon structure functions and then we have convoluted these 
nucleon structure functions with the spectral function in order to evaluate the nuclear structure
functions (Eq.\ref{spect_funct}). Using the nuclear structure functions, we have obtained the differential scattering cross 
sections  for the $\nu_l(\bar\nu_l)-A,\;\;l=e,\mu$ DIS process (Eq.\ref{xsecsf}). 
% The TMC and HT effects have already been discussed in detail in our recent work~\cite{Ansari:2020xne}. 

The calculations are performed in the four flavor MS-bar scheme. We have considered two cases: {\bf (i)} In the case of massless leptons ($e^\pm,~\mu^\pm$), all the four quarks, i.e., $u,~d$, $s$ and $c$ are treated to be massless while {\bf (ii)} For massive lepton ($\tau^\pm$), light quarks $u,~d$ and $s$ are treated as massless but charm quark $c$ as a massive object for which we define 
\begin{equation}\label{nf4_charm}
 F_{iA}(x,Q^2)=F_{iA}^{n_f=4}(x,Q^2)=\underbrace{F_{iA}^{n_f=3}(x,Q^2)}_{\textrm{for massless($u,d,s$) quarks}}+\underbrace{F_{iA}^{n_f=1}(x,Q^2)}_{\textrm{for massive charm quark}}.
\end{equation}
\begin{figure}
\begin{center}
\includegraphics[height=4. cm, width=3. cm]{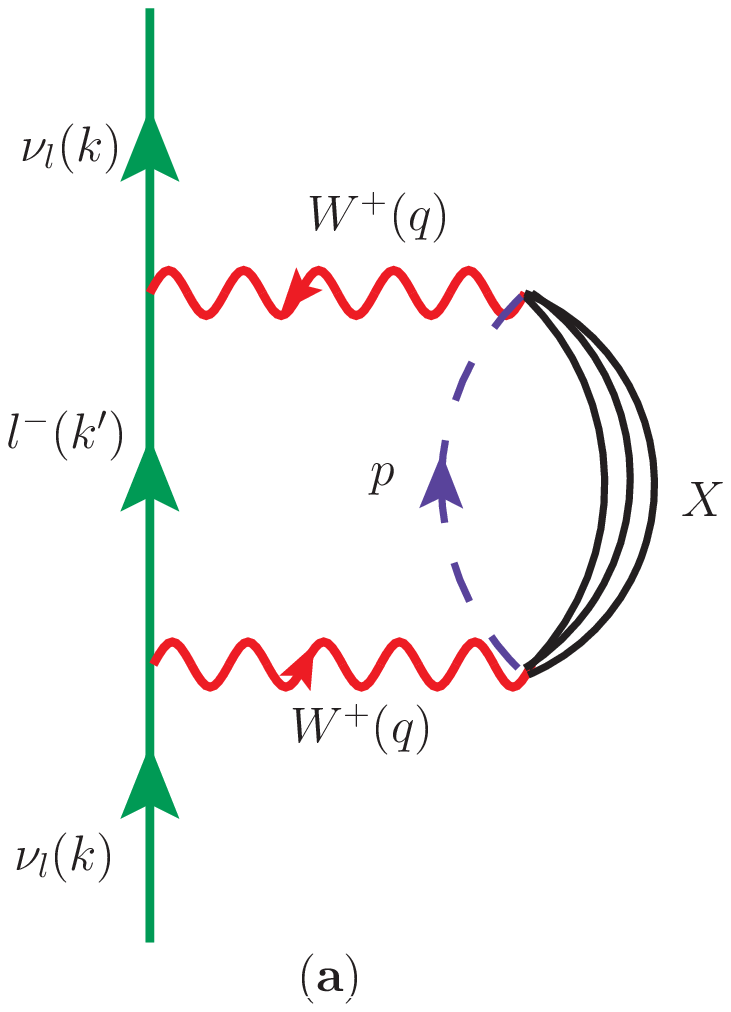}
 \includegraphics[height=4. cm, width=9 cm]{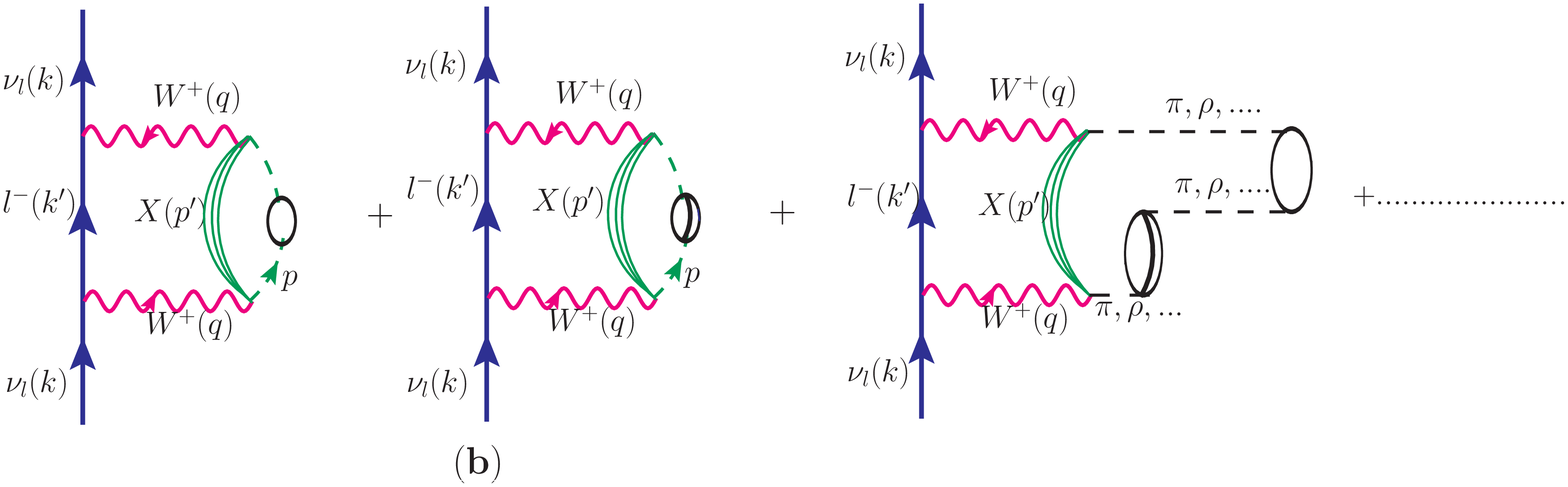}
 \end{center}
 \caption{Neutrino self energy diagram accounting for neutrino-meson DIS (a) the bound nucleon propagator is substituted with a meson($\pi$ or $\rho$) propagator with momentum $p$
 represented here by dashed line
 (b) by including particle-hole $(1p–1h)$, delta-hole $(1\Delta–1h)$,
 $1p1h-1\Delta1h$, etc. interactions.}
 \label{n_self-1}
\end{figure}
The nucleons which are bound inside the nucleus may interact with each other via meson exchange such as $\pi,~\rho,$ etc. This mesonic effect has been incorporated and is discussed in the next sub-subsection~\ref{mes}.
 \subsubsection{Effect of mesonic cloud contribution}\label{mes}
 Associated with each bound nucleon there are virtual mesons (pion, rho meson, etc.) and because of the strong
 attractive nature of the nucleon-nucleon interaction, the probability of a $W$-boson interaction with the mesonic cloud becomes high. We have discussed the $\pi$ and $\rho$ meson contributions in our earlier works~\cite{Haider:2016zrk,Zaidi:2019mfd,Zaidi:2019asc,Haider:2015vea,Haider:2016tev}. The contribution from heavier mesons like $\omega$ meson are expected to be very small due to their significantly higher masses. The mesonic contribution is significant in the intermediate region of $x~ (0.2 < x < 0.6)$. For the medium nuclei like $^{4}He$, $^{12}C$, etc., mesonic contribution is small~\cite{Haider:2011kvi} while 
 it becomes pronounced in heavier nuclear  targets~\cite{Haider:2012nf} such as $^{56}Fe$, $^{208}Pb$, etc. 
To incorporate the mesonic effect we draw a diagram similar to Fig.\ref{wself_energy}, but here a nucleon propagator is replaced by a meson propagator. 
The meson propagator corresponds to the virtual mesons arising due to the nuclear medium effects and can arise through the 
 particle-hole $(1p–1h)$, delta-hole $(1\Delta–1h)$, $2p-2h$, etc. interactions as shown in Fig.\ref{n_self-1}~\cite{FernandezdeCordoba:1991wf}. 
 
 The mesonic structure functions $F_{{i A,a}}(x_a,Q^2), ~~(i=1,2;a=\pi,\rho)$ are obtained as~\cite{Zaidi:2019asc}:
 \begin{eqnarray} 
\label{pion_f21}
F_{{i A,a}}(x_a,Q^2)  &=&  -6 \kappa \int \, d^3 r  \int \frac{d^4 p}{(2 \pi)^4} 
        \theta (p^0) ~\delta I m D_a (p) \;2m_a~\;f_{ia}(x_a),
\end{eqnarray}
 where
 \begin{eqnarray} 
\label{F2rho_wk}
f_{1a}(x_a) &=&  A M_N \left[\frac{F_{1a}(x_a)}{m_a}~+~\frac{{|{\bf p}|^2~-~(p^{z})^2}}{2(p^0~q^0~-~p^z~q^z)}
\frac{F_{2a}(x_a)}{m_a}\right],~~~~~
\end{eqnarray}
 \begin{eqnarray}
 f_{2a}(x_a) &=&\left[\frac{Q^2}{(q^z)^2}\left( \frac{|{\bf p}|^2~-~(p^{z})^2}{2m_a^2}\right)  +  \frac{(p^0~-~p^z~\gamma)^2}{m_a^2}\times\left(\frac{p^z~Q^2}{(p^0~-~p^z~\gamma) q^0 q^z}~+~1\right)^2\right] \nonumber\\
 &\times&\left(\frac{m_a}{p^0~-~p^z~\gamma}\right)~F_{{2a}}(x_a).~~~~~~
 \end{eqnarray}
In Eqs.~\ref{pion_f21} and \ref{F2rho_wk}, $\kappa=1(2)$ for pion(rho meson), $\nu=\frac{q_0(\gamma p^z-p^0)}{m_a}$, $x_a=-\frac{Q^2}{2p \cdot q}$, $m_a$ is the mass of the meson($\pi$ or $\rho$). $D_a(p)$ is the meson($\pi$ or $\rho$) propagator in the nuclear medium and is written as 
 \begin{equation}\label{dpi}
D_a (p) = [ p_0^2 - {\bf {p}}\,^{2} - m_a^2 - \Pi_{a} (p_0, {\bf p}) ]^{- 1}\,,
\end{equation}
with
\begin{equation}\label{pionSelfenergy}
\Pi_a(p_0, {\bf p})=\frac{f^2}{m_\pi^2}\;\frac{C_\rho\;F^2_a(p){\bf {p}}\,^{2}\Pi^*}{1-{f^2\over m_\pi^2} V'_j\Pi^*}\,,
\end{equation}
where $C_\rho=1(3.94)$ for pion(rho meson). $F_a(p)={(\Lambda_a^2-m_a^2) \over (\Lambda_a^2 - p^2)}$ is the $\pi NN$ or $\rho NN$ form 
factor, 
% $p^2=p_0^2~-~{\bf p}^2$, 
$\Lambda_a$=1~$GeV$~\cite{SajjadAthar:2009cr,Haider:2015vea} and $f=1.01$.  $V_j'$ is the longitudinal(transverse)
part of the spin-isospin interaction for pion(rho meson), and $\Pi^*$ is the irreducible meson self energy that contains the contribution of particle-hole and delta-hole excitations. 
 For the pions, we have used the PDFs parameterization given by Gluck et al.\cite{Gluck:1991ey} and for the $\rho$ mesons same PDFs as for the pions have been used as there is no available explicit parameterization for the $\rho-$meson PDFs in the literature.
\subsubsection{Shadowing and Antishadowing effects}
\label{sec_shad}
In this work the shadowing and antishadowing effects are taken into consideration following the works of Kulagin and Petti~\cite{Kulagin:2007ju,Kulagin:2004ie} who have used the original Glauber-Gribov multiple scattering theory. For example, 
with these effects, the nuclear structure function is given by
\begin{equation}
 F_{iA,shd}(x,Q^2)= \delta R(x,Q^2) \times F_{iN}^{WI}(x,Q^2)\;\;\; i=1-3,
 \label{shdw11}
\end{equation}
where the expression for $\delta R(x,Q^2)$ used in the present numerical calculations is given in Ref.~\cite{Kulagin:2004ie}.

Now, using the present formalism, we have presented the results for the weak structure functions and scattering 
cross sections for both the free nucleon and nuclear targets in the next section. 
\begin{figure}
\centering
 \includegraphics[height=12 cm, width=0.95\textwidth]{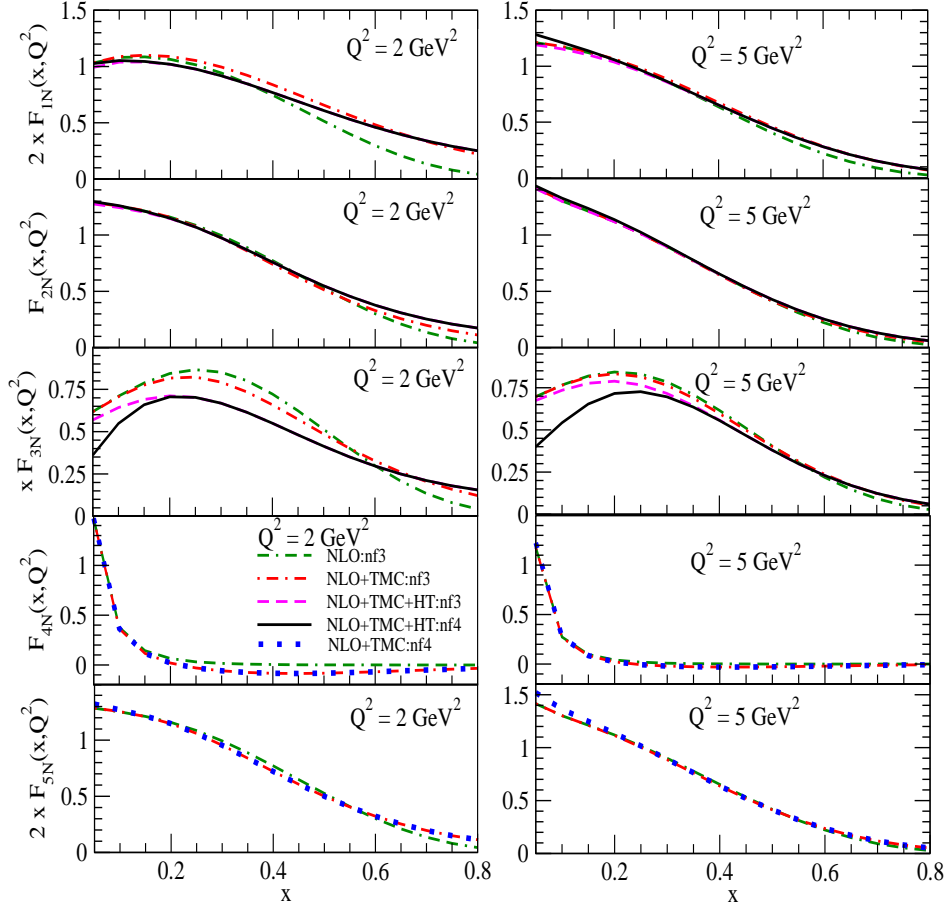}
 \caption{ Results for the free nucleon weak structure functions $F_{iN}(x,Q^2)$;$(i=1-5)$(Top to Bottom)
  at the different values of $Q^2$= 2 GeV$^2$(Left) and 
 5 GeV$^2$ (Right) are shown. These results are obtained at NLO by using MMHT nucleon PDFs
 parameterization~\cite{Harland-Lang:2014zoa}. The results are shown without the TMC effect (double dashed-dotted line), 
 with the TMC effect in the 3-flavor(nf3) scheme (dashed-dotted line) as well as four flavor(nf4) scheme(dotted line), with TMC and HT effects in the 3-flavor(nf3) scheme (dashed line) as well as four flavor(nf4) scheme(solid line).}
\label{fig1}
  \end{figure}
      \begin{figure}
 \includegraphics[height=14 cm, width=0.9\textwidth]{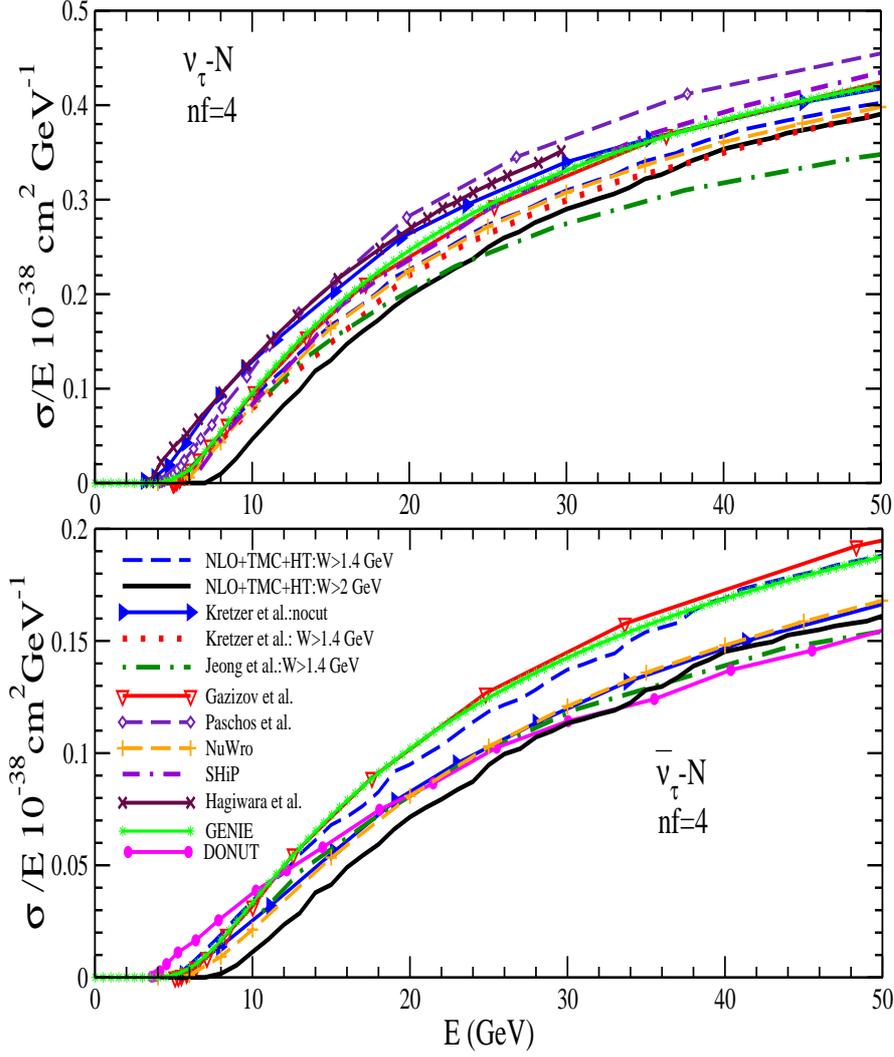}
 \caption{$\frac{\sigma}{E}$ vs $E$ for $\nu/{\bar\nu}_\tau$ interactions on a free nucleon target, with a center of mass energy ($W$) cut of 1.4 GeV(dashed line) and 2 GeV(solid line), for tau type neutrinos(Top panel)
 and antineutrinos(Bottom panel) with the TMC~\cite{Kretzer:2003iu} and higher twist~\cite{Dasgupta:1996hh} effects. These results are  compared with the results of various theoretical models available in the literature~\cite{Kretzer:2002fr,Jeong:2010nt,Hagiwara:2003di,Conrad:2010mh,Paschos:2001np,Gazizov:2016,Li:2017dbe,Anelli:2015pba} as well as with the Monte Carlo generators GENIE~\cite{GENIE} and NuWro\cite{NuWro}.}
 \label{fig:sig_comp}
\end{figure}
 \section{Results and Discussion}
 First, we present the results for the free nucleon weak structure functions $F_{iN}(x,Q^2)$;$(i=1-5)$, and using them the results for the total scattering cross section $\sigma(E_\nu)$ are obtained and presented for $\frac{\sigma(E_\nu)}{E_\nu}$ vs $E_\nu$ in the limit $m_l \ne 0$.
 
Then we have presented the results for the weak nuclear structure functions $F_{iA}(x,Q^2)$;$(i=1-3)$ in the limit $m_l = 0$,
 when the calculations are performed using the spectral function (SF) only. 
 Furthermore, the contribution from the meson clouds as well as the shadowing effect 
 are taken into account and this corresponds to the full model (Total) results. 
 The expression of total nuclear structure functions in the full theoretical model is given by
\begin{eqnarray}\label{f1f2_tot}
 F_{iA}(x,Q^2)=F_{iA,N}(x,Q^2)+F_{iA,\pi}(x,Q^2)+F_{iA,\rho}(x,Q^2)
 +F_{iA,shd}(x,Q^2),\;\;\;\;\;
\end{eqnarray}
where $i=1,2$. $F_{iA,N}(x,Q^2)$ are the weak nuclear structure functions which have contributions from the spectral function only, $F_{iA,\pi/\rho}(x,Q^2)$ take into 
account mesonic contributions. $F_{iA,shd}(x,Q^2)$ has contribution from the 
shadowing effect.

  \begin{figure}
 \includegraphics[height=10 cm,width=0.95\textwidth]{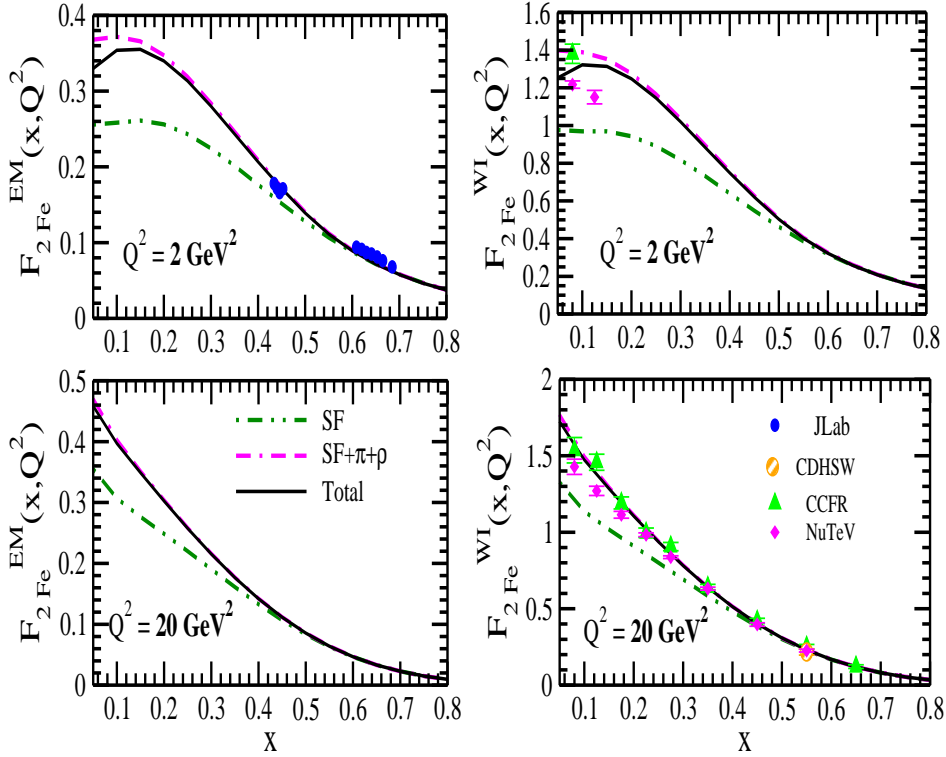}
 \caption{Results of EM(Left panel) and Weak(Right panel) nuclear structure functions in $^{56}Fe$(isoscalar) obtained 
 using spectral function(dashed double-dotted line), including mesonic 
contribution(double dashed-dotted line), full model(solid line). These results are presented for different $Q^{2}$ and are compared with the available 
 experimental data of JLab(solid circle)~\cite{Mamyan:2012th},
 CDHSW(semi solid circle)~\cite{Berge:1989hr}, CCFR(triangle up)~\cite{Oltman:1992pq}, NuTeV(diamond)~\cite{Tzanov:2005kr}.}
 \label{fig2}
\end{figure}

In this model, the full expression
for the parity violating weak nuclear structure function $F_{3A}(x,Q^2)$ is given by,
\begin{eqnarray}\label{f3_tot}
 F_{3A}(x,Q^2)= F_{3A,N}(x,Q^2) + F_{3A,shd}(x,Q^2).
\end{eqnarray}
Notice that this structure function has no mesonic contribution. The contribution to the nucleon structure function mainly comes from the valence quarks distributions. 
For $F_{3A,shd}(x,Q^2)$ similar definition has been used as given
in Eq.(\ref{shdw11}) following the works of Kulagin et al.~\cite{Kulagin:2004ie}.
\begin{figure}
\begin{center}
 \includegraphics[height= 8 cm , width= 0.95\textwidth]{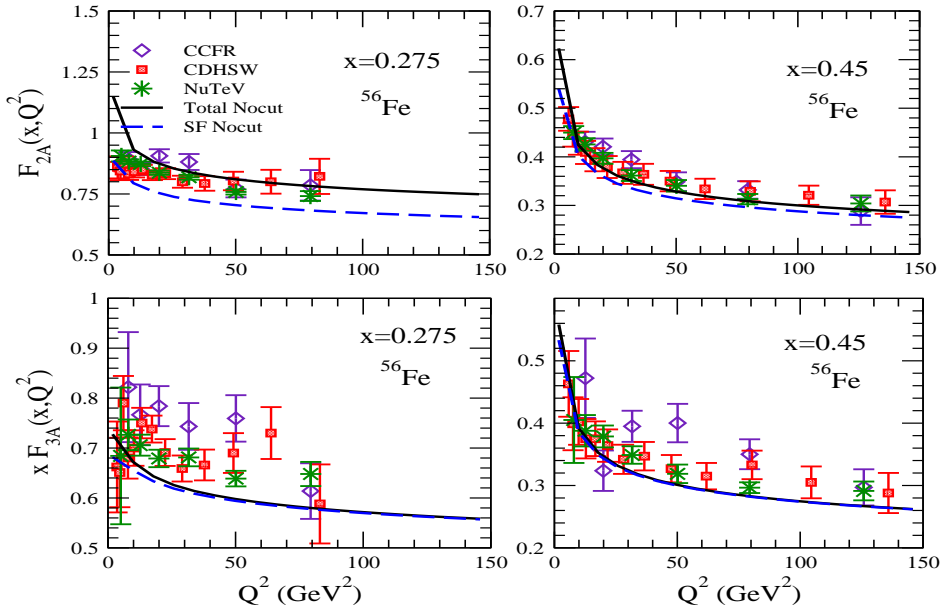}
\end{center}
\caption{Results for $ F_{2A}^{\nu+\bar\nu}(x,Q^2)$ (top panel) and $x F_{3A}^{\nu+\bar\nu}(x,Q^2)$ (bottom panel) vs $Q^2$ are shown at $x=0.275$ and 0.45 in $^{56}$Fe  . The results are obtained with the spectral function only (dashed line) and with the full model (solid line) at NNLO and are compared with the results of the available 
experimental data~\cite{Berge:1989hr,Oltman:1992pq,Tzanov:2005kr}.
Iron is treated as an isoscalar nuclear target.}
\label{fig7}
\end{figure}
\subsection{Nucleon Structure Function and Cross Section}
  In Fig.~\ref{fig1}, the results for the free nucleon weak structure functions $2xF_{1N}(x,Q^2)$, $F_{2N}(x,Q^2)$, $xF_{3N}(x,Q^2)$, $F_{4N}(x,Q^2)$ and $2xF_{5N}(x,Q^2)$ 
  (from the top to bottom) are shown at the two different values of $Q^2$ viz. $Q^2=2$ GeV$^2$ (left panel) and $Q^2=5$ GeV$^2$
  (right panel). The nucleon structure functions are presented at NLO 
   without the TMC effect (double dash-dotted line), with the TMC effect in 3-flavor(dash-dotted line:nf3) and 4-flavor(dotted line:nf4) schemes, with TMC and HT effects in 3-flavor(dashed line: nf3) and 4-flavor(solid line:nf4) schemes. From the figure, it may be noticed that the TMC effect is dominant in the region of high $x$ and low $Q^2$ and it becomes small at low $x$ and high $Q^2$. Quantitatively, the TMC effect is found to be different in $F_{2N}(x,Q^2)$ from $F_{1N}(x,Q^2)$ while 
  the TMC effect in $F_{5N}(x,Q^2)$ is similar to the effect in $F_{2N}(x,Q^2)$. However, in the case of $F_{4N}(x,Q^2)$ the whole contribution, arises in the leading order due to the TMC effect at mid and high $x$. $F_{4N}(x,Q^2)$ contributes to the cross section due to the non-vanishing lepton mass when it is large, and  contributes only in the region of $x \le 0.2$. We find that at NLO, $F_{4N}(x,Q^2)$ becomes almost negligible in the region of $x>0.2$ when TMC effect is not incorporated but with the inclusion of TMC effect a nonzero though small contribution in the region of high $x$ and low $Q^2$ has been found. The difference in the results of free nucleon structure functions $F_{iN} (x, Q^2);~~ (i = 1-5)$ evaluated at NLO with and without the TMC effect at $x=0.3$ is $5\%(3\%)$ in $2xF_{1N} (x, Q^2)$, $2\%(<1\%)$ in $F_{2N} (x, Q^2)$, $7\%(\sim 3\%)$ in $xF_{3N} (x, Q^2)$ and $4\%(\sim 2\%)$ in $2xF_{5N} (x, Q^2)$ for $Q^2=2(5)$ GeV$^2$.
  
 For the first three structure functions ($F_{iN}(x,Q^2);~~i=1-3$), the HT effect has also been studied. This is found to be comparatively small in $F_{1N}(x,Q^2)$ and $F_{2N}(x,Q^2)$ than in $F_{3N}(x,Q^2)$. Due to higher twist(HT) corrections, we have observed a decrease in the value of  $F_{3N}(x,Q^2)$, which becomes small with the increase in $Q^2$. 
   To show the effect of massive charm on the nucleon structure functions, we have compared the results obtained with the TMC and HT effects for the three flavor of massless quarks ($n_f=3$) with the results when additional contribution from the massive charm quark ($n_f=4$) have also been considered. It is found that
   massive charm effect is almost negligible in the kinematic region of low $Q^2$ and high $x$ while it increases with the increase in $Q^2$ and is appreciable at low $x$. 
   \begin{figure}
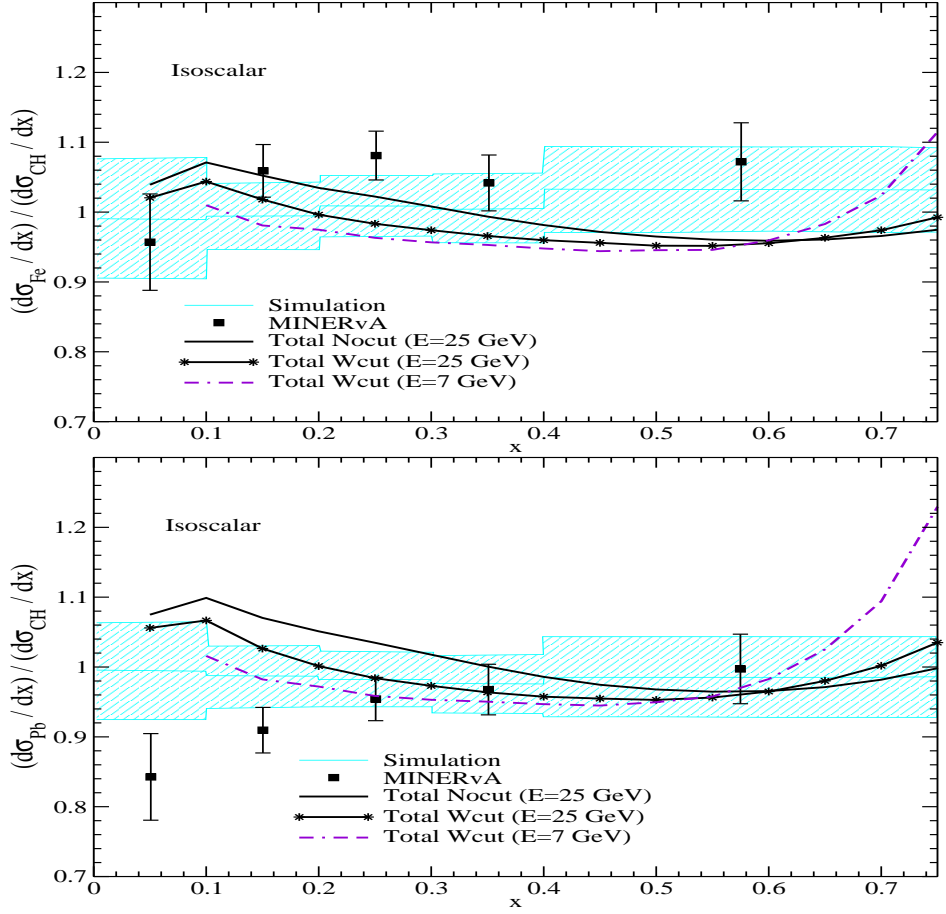

\begin{center}
    \includegraphics[height= 6 cm , width= 0.95\textwidth]{dsigma_fetoch_isoscalar.eps}\\
    \includegraphics[height= 6 cm , width= 0.95\textwidth]{dsigma_pbtoch_isoscalar.eps}
\end{center}
 \caption{$\frac{d\sigma_{A}/dx}{d\sigma_{CH}/dx}~(A=$ $^{56}$Fe, $^{208}$Pb) vs $x$ for incoming neutrino beam of energies $E=7$ GeV and $25$ GeV. The numerical results are 
 obtained with the full model (solid line: $E=25$ GeV, solid line with star: $E=25$ GeV
 and double dash-dotted line: $E=7$ GeV) at NNLO and are compared with the simulated results~\cite{Mousseau:2016snl}. The solid squares are the experimental points of MINERvA~\cite{Mousseau:2016snl}.
 The results are shown for the isoscalar nuclear targets.}
\label{fig9}
\end{figure}

 In Fig.~\ref{fig:sig_comp}, we have shown our results for the total scattering cross section $\sigma/E_{\nu}$ vs $E_{\nu}$, for the DIS of $\nu_\tau$ and ${\bar\nu}_\tau$ from the nucleons and compared them  with the results of the different models available in the literature like that of Pashchos et al.~\cite{Paschos:2001np} (dashed line with diamond), Kretzer et al.~\cite{Kretzer:2002fr} (solid line with right triangle without a cut on $W$; dotted line with a cut of $W > 1.4$ GeV), Jeong et al.\cite{Jeong:2010nt} (dash-dotted line), Gazizov et al.\cite{Gazizov:2016} (solid line with down triangle), Hagiwara et al.\cite{Hagiwara:2003di} (solid line with cross symbol), Anelli et al.\cite{Anelli:2015pba} (double dash-dotted line) and Li et al.~\cite{Li:2017dbe} (solid line with circles) as well as with the predictions made by the Monte Carlo generator GENIE~\cite{GENIE} and NuWro~\cite{NuWro}. The results are presented for the two cases of cut on the center of mass energy(W), taken to be W=1.4 GeV(shown by dashed line) and 2 GeV(shown by the solid line). 
 The results are presented by incorporating the target mass correction and higher twist effects at NLO in the four flavor scheme. 
 Our results with a cut of $W= 1.4~ GeV$ (shown by dashed line) is in good agreement with the result of Kretzer et al.~\cite{Kretzer:2002fr} (shown by the dotted line) while there are significant differences from the result of Jeong et al.~\cite{Jeong:2010nt} (shown by the dash-dotted line). Notice that the results of the total scattering cross section with the same CoM energy cut reported by Kretzer and Reno~\cite{Kretzer:2002fr} and Jeong and Reno~\cite{Jeong:2010nt} are  also different with each other. The difference is mainly due to the choice of lower cuts on $Q^2$ in the evaluation of PDFs. It is important to point out that the results given by the different models~\cite{Kretzer:2002fr,Jeong:2010nt,Hagiwara:2003di,Conrad:2010mh,Paschos:2001np,Gazizov:2016,Li:2017dbe,Anelli:2015pba} have significant differences due to their choice of different kinematic conditions.  Furthermore, we have observed that the
dependence of the cross section on the  
  effect of CoM energy cut is more pronounced in the case of $\bar\nu_\tau-N$ DIS than in $\nu_\tau-N$ DIS process. 
  Moreover, one may also notice that the total scattering cross section is quite sensitive to the kinematic cut on the  CoM energy. It implies that a suitable choice of the CoM energy cut as well as the four momentum transfer square ($Q^2$) cut to define the deep inelastic region and use them to calculate the nucleon structure functions, differential and total 
  scattering cross sections is quite important. These kinematic considerations should be kept in mind while comparing the predictions of 
  the cross sections in various theoretical models.
 
\begin{figure}
\begin{center}
\includegraphics[height=8 cm, width=0.95\textwidth]{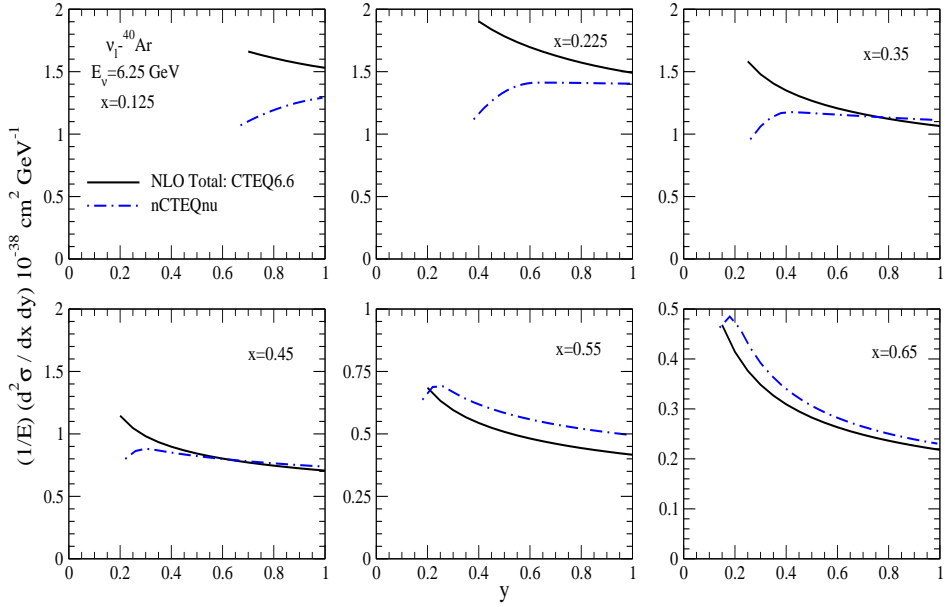}
\caption{Predictions for the differential scattering cross section vs y, at the different values of x for 6.25 GeV $\nu$-Ar treated as an isoscalar target. The results are obtained with a $Q^2 \ge 1.0 ~GeV^2$ cut by the Aligarh-Valencia model using CTEQ 6.6 nucleon PDFs~\cite{Nadolsky:2008zw} at NLO in the MS-bar scheme (solid line). The nCTEQnu nuclear PDFs~\cite{private} based prediction is the blue dash-dotted line. 
%The lower-y downward curve of the prediction at a given x corresponds to the extrapolation of the nCTEQnu global fit below $Q_0^2$ = 1.69 $GeV^2$.
}
\label{fig:10}
\end{center}
\end{figure}

\subsection{Nuclear Structure Function and Cross Section}
In Fig.~\ref{fig2}, we present the results for the weak nuclear structure function $F_{2A}^{Weak}(x,Q^2)$ and compare them with the electromagnetic structure function $F_{2A}^{EM}(x,Q^2)$ as a function of $x$ for the two
 different values of $Q^2$ viz. 2 and 20 $GeV^2$. 
 In this figure, we show the curves for $F_{2A}^{EM,Weak}(x,Q^2)$ obtained using the spectral
 function only, also including the mesonic contribution, 
 and finally using the full model which also includes shadowing. 
 The use of nuclear spectral function leads to a reduction of $\sim 8\%$ at $x=0.1$; $\sim 18\%$ at $x=0.4$; $\sim 3\%$ at $x=0.7$
 in $F_{2A}^{EM}(x,Q^2)$ as well as in $F_{2A}^{Weak}(x,Q^2)$ nuclear structure functions as compared to the free nucleon case. 
 The inclusion of mesonic contributions from pion and rho mesons leads to an enhancement in these structure functions at low and medium values of $x$. For example, 
  the enhancement is $\sim 30\%$ at $x$=0.1; $\sim 15\%$ at $x$=0.4; $\sim 0.3\%$ at $x$=0.7. The inclusion of shadowing effects further 
   reduces these structure functions and are effective in low region of $x~(~<~0.1)$. For example, 
  the reduction is $\sim 10\%$ at $x$=0.05 and $\sim 5\%$ at $x$=0.1.
 
 In Fig. \ref{fig7}, we present the results for the (anti)neutrino induced processes in $F_{2A}^{\nu+\bar\nu}(x,Q^2)$ 
and $x F_{3A}^{\nu+\bar\nu}(x,Q^2)$ vs $Q^2$ in $^{56}$Fe by treating it as an isoscalar nuclear target at the different values of $x$ using the full model at NNLO and compared them with the available
 experimental data from CCFR~\cite{Oltman:1992pq}, CDHSW~\cite{Berge:1989hr} and NuTeV~\cite{Tzanov:2005kr} experiments. 
 We find a good agreement between the theoretical results for $F_{2A}^{\nu+\bar\nu}(x,Q^2)$ and reasonable agreement for $F_{3A}^{\nu+\bar\nu}(x,Q^2)$ with the experimental data. It can be seen that the description of the nuclear medium effect in our model is able to reproduce the experimental results.

 Furthermore, in Fig.\ref{fig9}, we have presented the results for the ratio of the differential scattering cross sections, for the 
 different nuclear targets viz.  A= $^{56}$Fe and  $^{208}$Pb to the hydrocarbon(CH) target i.e.
 $\frac{d\sigma_{A}/dx}{d\sigma_{CH}/dx}$~(A= $^{56}$Fe, $^{208}$Pb) vs $x$ at the two different values of neutrino energies viz. $E_\nu=7$ GeV and $E_\nu=25$ GeV and compared them with the corresponding experimental data of MINERvA~\cite{Mousseau:2016snl}. It may be noticed that MINERvA's experimental data have large error bars 
 and the wide band around the simulation is due to the systematic errors which shows an uncertainty up to $\sim 20\%$~\cite{Mousseau:2016snl}. 

% For the future DUNE neutrino oscillation experiment~\cite{Abi:2018alz} Fig.~\ref{fig:10} shows predictions of both the full theoretical model and the nCTEQnu nPDFs for the differential cross sections with a 6.25 GeV neutrino beam on $^{40}Ar$. As $y$ decreases, the nPDF approach predicts lower cross sections than the theoretical approach.  For high-$x$ ($\gtrapprox$ 0.5), the nPDF approach and the theoretical approach predicts quite similar cross sections while for low-$x$ ($\lessapprox$ 0.3) the nPDF approach predicts a lower cross section than the theoretical approach.

Fig.~\ref{fig:10} shows predictions for the differential scattering cross section for $\nu_\mu$ induced reaction on $^{40}$Ar nuclear target at $E_{\nu_\mu}$=6.25 GeV. The numerical calculations are done by using the CTEQ6.6 PDFs parameterization~\cite{Nadolsky:2008zw}. Our theoretical results obtained using the full model are compared with the results of nCTEQnu nPDFs~\cite{private}. As $y$ decreases, the nPDF approach predicts lower cross sections than the theoretical approach. For high-$x$ ($\gtrapprox$ 0.5), the nPDF approach and the theoretical approach both predict quite similar cross sections while for the low-$x$ ($\lessapprox$ 0.3) region the nPDF approach predicts a lower cross section than our theoretical model.  

             \subsection{Nonisoscalarity Effect}
   For a nonisoscalar nuclear target like $^{56}Fe$ and $^{208}Pb$, where $(A-Z) > Z$, we obtain the normalized spectral function to the proton ($S_h^p$) and neutron ($S_h^n$) numbers (Eqs.\ref{conv_WAan}, \ref{spec_normp} and \ref{spec_normn}). For details please see  the discussion in Ref.\cite{Zaidi:2019mfd}.
%    and write the structure function of the nucleon as
%$$F_{iN}(x,Q^2)={F_{ip}(x,Q^2)+F_{in}(x,Q^2) \over 2};~~~i=1-3 $$
\begin{figure}
\begin{center}
 \includegraphics[height= 9 cm , width= 0.95\textwidth]{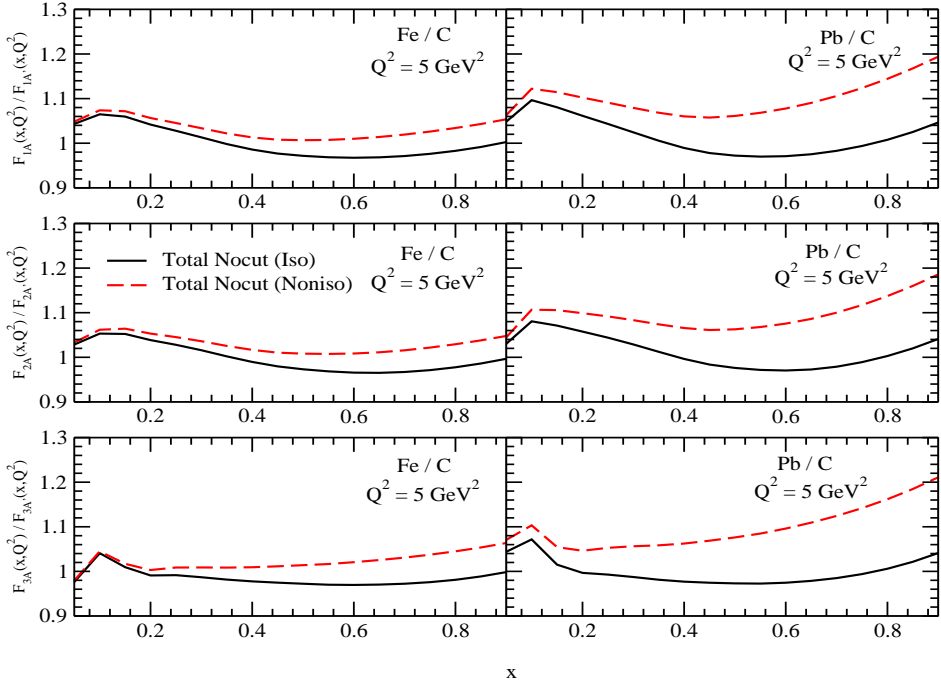}
\end{center}
\caption{$\frac{F_{iA}(x,Q^2)}{F_{iA'}(x,Q^2)};~(i=1-3$; $A=^{56}$Fe, $^{208}$Pb; $A'=^{12}$C) vs $x$ are shown at $Q^2=5$ GeV$^2$. The results are obtained using the full 
model at NNLO by 
treating $^{56}$Fe, $^{208}$Pb to be isoscalar (solid line) as well as nonisoscalar (dashed line) nuclear targets.}
\label{fig8}
\end{figure}
In Fig.\ref{fig8}, we have presented the results for the ratios of nuclear structure functions $\frac{F_{iA}(x,Q^2)}{F_{iA'}(x,Q^2)};~(i=1,2,3;~A= ^{56}Fe,~ ^{208}Pb~ \textrm{and}~ A'=^{12}C)$ vs $x$ at $Q^2=5$ GeV$^2$. The present numerical results are obtained by using the full model at NNLO and treating iron and lead to be isoscalar as well as nonisoscalar nuclear targets. We find that the ratios $\frac{F_{i Fe}(x,Q^2)}{F_{i C}(x,Q^2)}$ and $\frac{F_{i Pb}(x,Q^2)}{F_{i C}(x,Q^2)}$ deviate from unity in the 
  entire range of $x$ which implies that nuclear medium effects are $A$ dependent, and the medium effects become more pronounced with the increase in the nuclear mass number. This enhancement in the ratio is more at high $x$. 
             \section{Summary}
             \begin{enumerate}
\item The inclusion of perturbative and nonperturbative effects is quite important in the evaluation of nucleon structure functions. These effects are both $x$ and $Q^2$ dependent.
\item The difference in the results of all the free nucleon structure functions $F_{iN} (x, Q^2);~~(i = 1-5)$ evaluated at NLO with and without the TMC effect is non-negligible. In the case of $F_{4N}(x,Q^2)$ this difference is quite large when TMC effect is considered, specially at mid and high $x$. The higher twist(HT) effect has not much effect on $F_{1N} (x, Q^2)$ and $F_{2N} (x, Q^2)$, but affects $F_{3N} (x, Q^2)$. We find that the difference due to HT effect is somewhat larger for $F_{3N}(x,Q^2)$ at low $x$ and low $Q^2$. 
With the increase in $x$ the effect of HT increases.
\item The total cross section $\sigma$ for $\nu_\tau$ and ${\bar\nu}_\tau$ interactions on the free nucleon target shows significant  dependence on the model used to calculate the nucleon structure functions which implies that there are uncertainties even 
 at the level of free nucleon level and more work is needed. 
\item NME in $F_{2 A}^{EM}(x,Q^2)$  and  $F_{2 A}^{Weak}(x,Q^2)$ in iron nucleus are almost the same for $x$ i.e $x~>~0.3$,
 but are different in small $x$ region($x<0.3$). However, this difference is found to be very small for isoscalar nuclei, and is similar to the case 
  of free nucleon.
  \item The nuclear structure functions obtained with spectral function only are suppressed as compared to the free nucleon case in the entire region of $x$.
 Whereas, the inclusion of mesonic contributions results in an enhancement in the nuclear structure functions in the low and intermediate region of $x$. 
 Mesonic contributions are observed to be more pronounced with the increase in mass number and they decrease with the increase in $x$ and $Q^2$.
The results for the nuclear structure functions $F_{2A}^{Weak}(x,Q^2)$ and $F_{3A}^{Weak}(x,Q^2)$ with the full theoretical model show good agreement with the 
 experimental data of CCFR~\cite{Oltman:1992pq}, CDHSW~\cite{Berge:1989hr} and NuTeV~\cite{Tzanov:2005kr} especially at high $x$ and high $Q^2$.
 \item The present theoretical results for the ratio $\frac{d\sigma_A^{Weak}/dx}{d\sigma_{CH}^{Weak}/dx}~(A=^{56}Fe,~^{208}Pb)$ when
compared with the different phenomenological models and the recent MINERvA's experimental data on $\nu_l-A$ scattering, imply that a better understanding of nuclear medium effects is
required for the $\nu_l(\bar\nu_l)-$nucleus deep inelastic scattering process.  
 \item We find the non-isoscalarity effect to be non-negligible which increases with the non-isoscalarity $\delta(=\frac{N-Z}{A})$ and $x$.
\item Predictions are 
 also made for $\nu_\mu(\bar\nu_\mu)$ interactions on $^{40}Ar$ that may be useful in analyzing the experimental results of DUNE~\cite{Abi:2018alz}.
\end{enumerate}

\section*{Acknowledgment}   
F. Zaidi is thankful to the Council of Scientific \& Industrial Research (CSIR), India, for providing the research associate fellowship with 
award letter no. 09/112(0622)2K19 EMR-I. I. Ruiz Simo acknowledges support from Spanish Ministry of Science and ERDF under contract FIS2017-85053-C2-1P, and from  Junta de Andalucia through grant No FQM-225.

\end{document}